\documentclass[twocolumn]{pasj00}

\begin{document} 

\SetRunningHead{Wide-band Temporal Spectroscopy of Cygnus X-1 with Suzaku}{S. Yamada et al.}
\title{Evidence for a Cool Disk and Inhomogeneous Coronae \\ 
from Wide-band Temporal Spectroscopy of Cygnus X-1 with Suzaku} 

\author{%
Shin'ya \textsc{Yamada}\altaffilmark{1},
Kazuo \textsc{Makishima}\altaffilmark{2,1,4}, 
Chris \textsc{Done}\altaffilmark{3},
Shunsuke \textsc{Torii}\altaffilmark{2},
Hirofumi \textsc{Noda}\altaffilmark{2},
and Soki \textsc{Sakurai}\altaffilmark{2}
}

\altaffiltext{1}{
   Cosmic Radiation Laboratory, Institute of Physical and Chemical 
   Research (RIKEN), \\
   2-1 Hirosawa, Wako-shi, Saitama, 351-0198}   

\altaffiltext{2}{
   Department of Physics, University of Tokyo, 
   7-3-1, Hongo, Bunkyo-ku, Tokyo 113-0033}   


\altaffiltext{3}{Department of Physics, Durham University, South Road, Durham, DH1 3LE, UK}

\altaffiltext{4}{Research Center for the Early Universe, RESCEU, University of Tokyo, 7-3-1, Hongo, Bunkyo-ku, Tokyo 113-0033}

\email{yamada@crab.riken.jp}


\Received{2012, December, 12}
\Accepted{2013, April, 2}
\Published{2013, August, 25}

\maketitle

\begin{abstract}

Unified X-ray spectral and timing studies of Cygnus X-1 
in the low/hard and hard intermediate state were conducted
in a model-independent manner,  
using broadband Suzaku data acquired 
on 25 occasions from 2005 to 2009 
with a total exposure of $\sim 450$ ks. 
The unabsorbed 0.1--500 keV source luminosity changed 
over 0.8--2.8\% of the Eddington limit for 14.8 solar masses.  
Variations on short (1--2 seconds) 
and long (days to months) time scales 
require at least three separate components: 
a constant component localized below $\sim$2~keV, 
a broad soft one dominating in the 2--10 keV range, 
and a hard one mostly seen in 10--300 keV range.  
In view of the truncated disk/hot inner flow picture, 
these are respectively interpreted as emission 
from the truncated cool disk, 
a soft Compton component,  
and a hard Compton component. 
Long-term spectral evolution can be produced 
by the constant disk increasing in temperature and luminosity as the
truncation radius decreases. 
The soft Compton component likewise
increases, but the hard Compton does not, 
so that the spectrum in the hard intermediate state 
is dominated by the soft Compton component; 
on the other hand, 
the hard Compton component dominates the spectrum in the dim low/hard state, 
probably associated with a variable soft emission providing seed photons for the Comptonization. 

\end{abstract}


\section{Introduction}
\label{sec:intro}

X-rays have been an essential probe to study energetic astrophysical
phenomena, including mass accretion onto black holes in particular.
Starting with the first identification of Cygnus X-1 (hereafter Cyg
X-1) as a black hole binary in the early 1970's (e.g., \cite{Oda1971},
Tananbaum et al.~1972, \cite{Thorne1975}), the spectral as well as timing studies of black
hole binaries, including Cyg X-1, have established the presence of two distinct states: 
the high/soft state and the low/hard state (e.g., \cite{2006ARA&A..44...49R}, Done et al.~2007).

\begin{figure}[htbp]
    \begin{center}
      \includegraphics[width=0.4\textwidth]{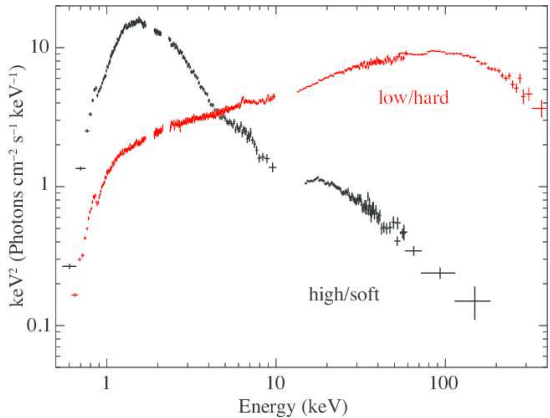}
            \end{center}
    \caption{Suzaku spectra of Cyg X-1 in the response-removed $\nu F_\nu$ form. The black one was obtained in the high/soft state on 2010 December 16. 
    The red one was taken in the low/hard state on 2005 October 5, 
    which is the same as used in Makishima et al. (2008).} 
    \label{suzakuspec}
\end{figure}

In the high/soft state, 
an optically thick and geometrically thin 
accretion disk \citep{1973A&A....24..337S} releases most of
the gravitational energy in locally thermal equilibrium radiation.  
The spectra in this state can be fairly well reproduced 
by multi-color blackbody radiation (Mitsuda et al.~1984 and Makishima et al.~1986) 
from such a ``standard'' accretion disk 
and a powerlaw emission with a photon index of $\sim 2.5$ 
of which the origin is commonly considered to be Compton scattering 
by a disk corona with a hybrid (thermal + non-thermal) electron distribution 
(e.g., \cite{Dotani1997}, Gierli\'{n}ski~et al.~1999, and \cite{2011Gou}), 
along with reflection features arising from disk regions 
illuminated by the powerlaw component. 
A typical high/soft state spectrum from Cyg X-1 is shown as black in figure \ref{suzakuspec}.

In the low/hard state, 
the spectrum is no longer dominated by the disk emission. 
Instead, 
most of the energy is released in hard X-rays via
un-saturated Comptonization by Maxwellian electrons 
(e.g., \cite{1979Natur.279..506S}). 
This Comptonizing region, conventionally
called a ``corona'', is hot (with a typical temperature of $\sim$ 100~keV, 
possibly with a much higher ion temperature, 
so most probably geometrically thick), and optically thin, in marked contrast
to the standard disk which is cool, geometrically thin and optically thick.  
However, some optically thick material is still present in this state: 
there is a weak thermal component seen in the soft X-ray bandpass 
which presumably provides seed photons for the Comptonization, 
and there is clear evidence for some reflected emission (iron lines and Compton
humps; e.g., Gierli\'{n}ski et al 1997; Gilfanov, Churazov \& Revnivtsev 1999; \cite{2001ApJ...547.1024D}, 
\cite{2004PThPS.155...99Z}, Ibragimov et al 2005). 
A typical low/hard state spectrum from Cyg X-1, highlighting the very different properties
of this state, is shown in red in figure \ref{suzakuspec}. 

The relative geometry of the disk and corona in the low/hard state is
still a matter of debate. One type of models assume that the corona
replaces the inner disk, with the flow making a transition to an
alternative, hot, geometrically thick, optically thin solution to the
accretion flow equations
(e.g., \cite{1976ApJ...204..187S}; \cite{1977ApJ...214..840I};
\cite{1995ApJ...444..231N}).  Exterior to this, the outer truncated
disk provides a source of seed photons for the Compton scattering from
the hot flow, and a site for the reflected emission. 
These truncated disk/hot inner flow models are 
successful in explaining the clear correlations seen in the data
(e.g Gilfanov, Churazov \& Revnivtsev 1999;
\cite{2004PThPS.155...99Z}, Done, Gierli\'{n}ski \& Kubota 2007). 
Alternatively, 
other geometries have also been proposed, 
the vertically separated sandwich ``disk-corona configuration''
(e.g., \cite{1977ApJ...218..247L}, \cite{Haardt1991}, \cite{1993Po}), 
the vertically offset ``lamppost'' model (e.g. Fabian et al 2012), 
and the vertically outflowing corona (Beloborodov et al. 1999).

A key difference between these models 
for the low/hard state is the location of the innermost disk radius. 
In the truncated disk model, 
it is assumed to be larger
than the innermost stable circular orbit, while in the other geometries the
disk is envisaged to extend down to the last stable orbit. However,
a number of spectral analyses (e.g.,
\cite{1997MNRAS.288..958G}; \cite{1998Zd}; \cite{2001ApJ...546.1027F})
and timing studies (Miyamoto \& Kitamoto 1989; Negoro et al. 1994;
\cite{1999ApJ...510..874N}; \cite{2000Gil}; \cite{2000Rev}; Remillard
\& McClintock 2006) 
were unable to unambiguously settle the issue. 
This is because the disk
emission is much weaker than the Comptonized emission in this state,
appearing only as a very subtle excess in the lowest energies as shown in figure~1.  
To constrain the disk emission requires the best possible constraints on
the broad band Comptonized emission. 


Suzaku, the fifth Japanese X-ray satellite, carries the X-ray Imaging
Spectrometer (XIS; Koyama et al. 2007) located at the foci of the
X-ray Telescope (XRT; Serlemitsos et al. 2007), and a non-imaging hard
X-ray instrument, the Hard X-ray Detector (HXD; Takahashi et al. 2007;
Kokubun et al. 2007; Yamada et al. 2011).  These two instruments
enable us to simultaneously measure a wide-band 
(typically 0.5--300 keV) spectrum of bright hard X-ray sources.  
With this capability,
Suzaku has observed Cyg X-1 25 times from 2005 to 2009 in the low/hard
state and hard intermediated state,  
over which the 1--10 keV flux varied by a factor of $\sim$ 3. 
The 0.5--300 keV spectra taken in the first observation
has been reproduced by Makishima et al.~2008 (Paper~I), 
invoking two (hard and soft) Comptonization components, 
%
a truncated disk, and reflection components (the ``double-Compton modeling'' itself was first applied to 
Cyg X-1 by Frontera et al.~2001 and to AGN by Magdziarz et al. 1998). 
Based on the ``double-Compton modeling'' 
and other observational facts as to Fe-K lines and the refection strength, 
they proposed that 
there is an overlapping region between the disk and corona, 
and changes in the coronal coverage fraction of the disk produce the fast variation. 
Although this view was confirmed in subsequent Suzaku observations 
(\cite{2011ApJ...728...13N}, Fabian et al 2012), 
these authors discussed several possible 
alternatives to the double-Compton view, 
including non-thermal Comptonization, jet 
emission, and complex ionized reflection. 

To disentangle such modeling degeneracy, 
we use the variability on 
different timescales and 
try to identify separate spectral components in a
model-independent way. 
This was already initiated by Torii et al.~2011 (Paper~II), 
who analyzed the HXD (PIN and GSO) data from the 25 observations 
for spectral and timing properties of the hard X-ray emission. 
At that time, they were not able to include the XIS data 
as these are often severely affected by photon
pileup. 
Because of this limited energy range, 
they could fit their data
by a single Comptonization component together with a simple reflection
model. 
Now that we have established a method to
correct XIS data for pile-up effects (Yamada et al. 2012), 
we can complement the work in Paper~II, 
by including all the 25 XIS data sets. 

Thus in this paper, 
we can use the entire broad bandpass of Suzaku 
to study the spectral evolution 
through the low/hard and hard-intermediate state. 

The distance to Cyg X-1 has been recently determined as
$D=1.86^{+0.12}_{-0.11}$ kpc, via a trigonometric parallax measurement
using Very Long Baseline Array \citep{Reid2011}.  This value is
consistent with an independent measurement using dust scattering halo
(Xiang et al. 2011).  Based on this distance, the black hole mass and its
inclination were derived as $14.8\pm1.0 M_{\odot}$ and
$27.1\pm0.8^{\circ}$ \citep{Orosz2011}, respectively.  We adopt these
values throughout the present paper. 
Unless otherwise stated, errors refer to 90\% confidence limits.

\begin{figure*}[htbp] 
    \begin{center}
      \includegraphics[width=0.98\textwidth]{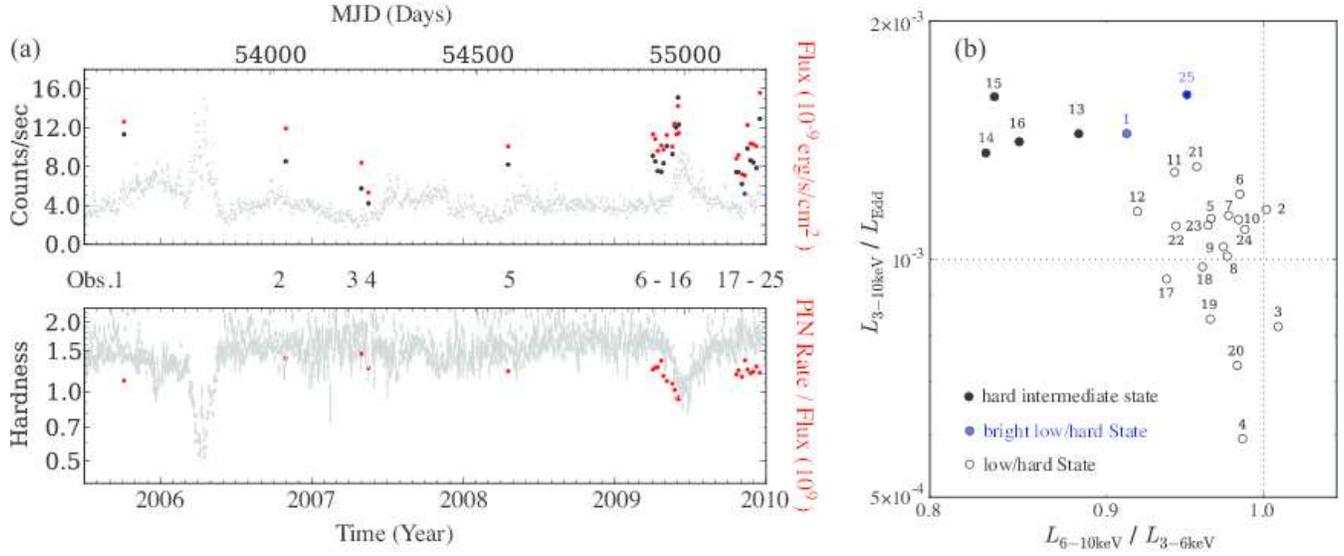}
            \end{center}
    \caption{(a) The 1.5--12 keV count-rate history of Cyg X-1 (gray), taken with the RXTE/ASM from 2005 July to 2010 January, which are scaled down to 25\%. The 0.5--10 keV flux of the XIS and the 15--20 keV count rate of PIN are superposed in red and black, respectively. The bottom panel shows the hardness ratios of 
    the 15--20 keV count rates of PIN to the 0.5--10 keV XIS fluxes ($\times 10^{9}$) (red dots), superposed on the 5--12 keV to 1.5--3 keV hardness of the ASM (gray). 
    (b) The 6--10 keV to 3--6 keV hardness ratio of the XIS plotted against the 3--10 keV intensity derived from the XIS normalized by the Eddington luminosity. All data are not corrected for the absorption.} 
    \label{alllc}
\end{figure*}

\begin{figure*}[htbp]
    \begin{center}
   \includegraphics[width=0.95\textwidth]{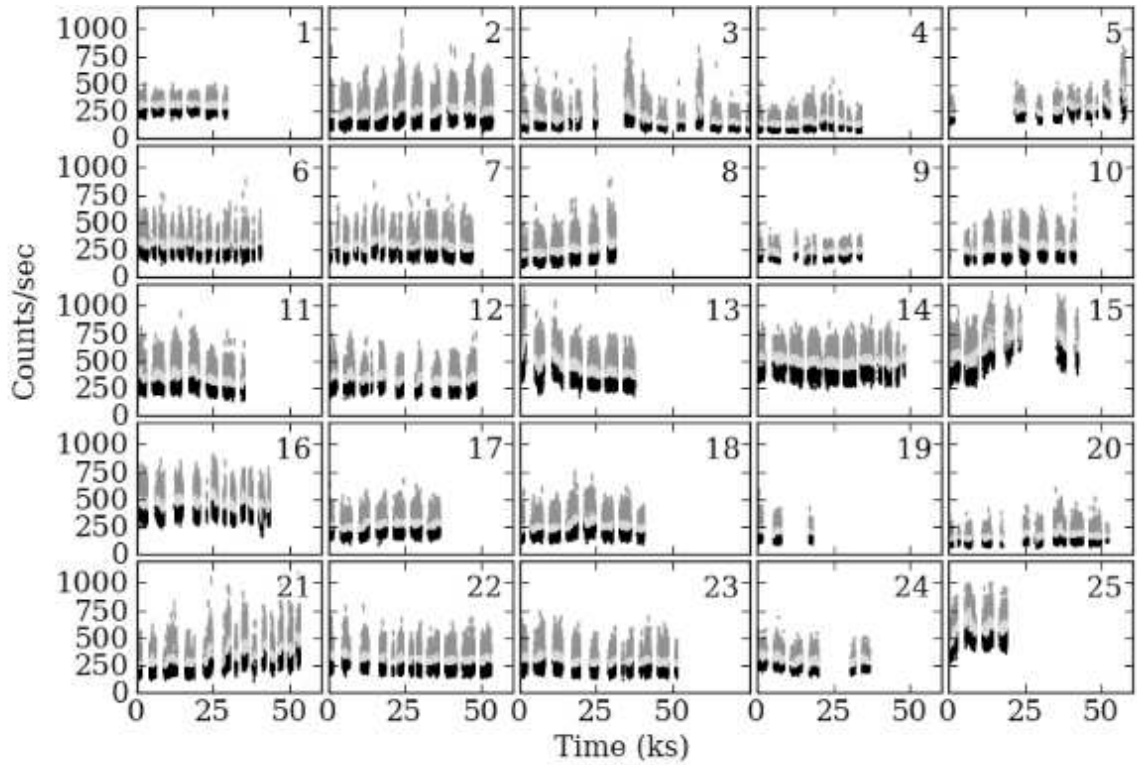}
            \end{center}     
    \caption{The 0.5--10 keV XIS3 light curves of Cyg X-1 for the 25 observations. The bin size is set at 1.0 s for 1/8 window options (Obs.~1 and 2) and timing mode (Obs.~5, 9, 24, and 25), while 2.0 s for the others. 
The data during dipping periods are removed. High-, intermediate-, and low-flux phases are shown in dark gray, light gray, and black, respectively.} 
    \label{hilolc}
\end{figure*}

\section{Observation and Data reduction}
\label{sec:obs}

\subsection{Observations}
\label{subsec:obs}

\begin{table*}[htbp]
 \caption{The log of Suzaku observations of Cyg X-1 taken from 2005 to 2009.}
 \label{obslog}
 \begin{center}
  \begin{tabular}{cccccccccc}
   \hline\hline
N  & Date (UT)  & T$^\dagger$ (ks) & Nom.$^*$ & Obs. ID & ASM$^*$  & Hard$^*$ & Orb. Phase$^\ddagger$  &  $F_{0.5-10}$$^\S$ & $L_{0.1-500}/L_{\rm{Edd}}$$^\|$  \\   [2mm] 
\hline
1 & 2005-10-05T04:51:39 & 37.2 & XIS   & 100036010	& 29.0 & 1.41 & 0.563--0.640 & 11.3  & 2.80(1.70)\\
2 & 2006-10-30T03:36:11 & 58.1 & HXD & 401059010	& 21.2 & 1.41 & 0.199--0.319 & 8.5 & 1.83(1.56)\\
3 & 2007-04-30T19:35:34 & 83.6 & HXD & 402072010	& 12.9 & 1.82 & 0.819--0.992 & 5.7 & 1.28(1.18)\\
4 & 2007-05-17T19:41:23 & 71.5 & HXD & 402072020	& 14.0 & 1.59 & 0.856--0.003 & 4.2  & 0.85(0.77)\\
5 & 2008-04-18T16:21:38 & 64.1 & HXD & 403065010	& 17.2 & 2.00 & 0.011--0.144 & 8.2  & 1.57(1.36)\\

6 & 2009-04-03T01:17:23 & 44.2 & HXD & 404075010	& 22.0 & 1.54 & 0.401--0.492 & 9.1 &1.87(1.52)\\
7 & 2009-04-08T06:09:03 & 48.9 & HXD & 404075020	& 21.7 & 1.61 & 0.330--0.431 & 8.5 & 1.70(1.45)\\
8 & 2009-04-14T18:21:40 & 33.0 & HXD & 404075030	& 19.7 & 1.54 & 0.492--0.561 & 7.5 & 1.53(1.29)\\
9 & 2009-04-23T04:01:10 & 39.6 & HXD & 404075040	& 20.0 & 1.54 & 0.993--0.075 & 7.4 & 1.53(1.36)\\
10 & 2009-04-28T17:02:23 & 43.9 & HXD & 404075050	& 22.6 & 1.39 & 0.983--0.073 & 8.3 & 1.58(1.36)\\

11 & 2009-05-06T16:48:59 & 37.6 & HXD & 404075060	& 21.0 & 1.61 & 0.410--0.487 & 10.1 & 1.85(1.45)\\
12 & 2009-05-20T00:34:56 & 50.1 & HXD & 404075080	& 23.9 & 1.12 & 0.789--0.892 & 9.3 &  1.63(1.23)\\
13 & 2009-05-25T08:35:49 & 40.4 & HXD & 404075090	& 31.8 & 1.05 & 0.741--0.825 & 12.2 & 2.00(1.44)\\
14 & 2009-05-29T11:53:46 & 48.6 & HXD & 404075100	& 33.7 & 1.12 & 0.480--0.581 & 12.0 & 1.97(1.22)\\
15 & 2009-06-02T11:33:13 & 43.7 & HXD & 404075110	& 45.0 & 1.05 & 0.192--0.282 & 15.9 & 2.28(1.43)\\

16 & 2009-06-04T19:42:00 & 44.5 & HXD & 404075120	& 35.0 & 1.12 & 0.610--0.702 & 12.3 & 2.05(1.29)\\
17 & 2009-10-21T09:03:11 & 40.0 & HXD & 404075130	& 18.9 & 1.43 & 0.353--0.435 & 7.4 & 1.40(1.11)\\
18 & 2009-10-26T06:21:56 & 41.7 & HXD & 404075140	& 18.9 & 1.59 & 0.226--0.312 & 7.4 & 1.41(1.20) \\
19 & 2009-11-03T21:28:15 & 44.6 & HXD & 404075150	& 14.0 & 1.69 & 0.767--0.859 & 6.2 & 1.70(0.98) \\
20 & 2009-11-10T19:39:56 & 53.4 & HXD & 404075160	& 17.4 & 1.89 & 0.003--0.114 & 5.1 & 1.05(0.98) \\

21 & 2009-11-17T06:51:08 & 56.7 & HXD & 404075170	& 24.4 & 1.75 & 0.158--0.275 & 9.8 & 1.81(1.56)\\
22 & 2009-11-24T12:20:29 & 53.5 & HXD & 404075180	& 22.4 & 1.59 & 0.449--0.559 & 8.6 & 1.63(1.33)\\
23 & 2009-12-01T07:00:50 & 51.7 & HXD & 404075190	& 25.0 & 1.61 & 0.659--0.766 & 8.4 & 1.65(1.36)\\
24 & 2009-12-08T15:32:11 & 39.5 & HXD & 404075200	& 22.2 & 1.75 & 0.973--0.054 & 7.8 &   1.56(1.38) \\
25 & 2009-12-17T01:29:11 & 42.2 & HXD & 404075070	& 40.9 & 1.59 & 0.475--0.563 & 12.9 & 2.28(1.87) \\
\hline\hline
\end{tabular}
\end{center}
\begin{itemize}
\setlength{\parskip}{0cm} %
\setlength{\itemsep}{0cm} %
\item[$^*$] ``Nom'' is the nominal pointing mode; ``ASM'' is the 1.5--12 keV count rate obtained with the RXTE All Sky Monitor; ``Hard'' is the hardness ratio of 5--12 keV to 1.5--3 keV ASM count rates. 
\item[$^\dagger$] From the start to the end of the observation. 
\item[$^\ddagger$] Using the orbital period of 5.599829 days, with phase 0 (the epoch of a superior conjunction of the black hole) being MJD$41874.207$ \citep{Brocksopp1999}. 
\item[$^\S$] The absorbed flux in $0.5$--$10$ keV in units of 10$^{-9}$ erg s$^{-1}$ cm$^{-2}$.
\item[$^\|$] The unabsorbed luminosity in $0.1$--$500$ keV divided by $L_{\rm{Edd}}$ and absorbed ones in the parentheses
\end{itemize}
\end{table*}


Like Paper~II, the present paper deals with the 25 Suzaku data sets of
Cyg X-1, of which basic information is listed in table \ref{obslog}.
The grey points in figure \ref{alllc}a presents the long term
count-rate and hardness-ratio histories of Cyg X-1 from the RXTE
All Sky Monitor (ASM) public data. 
The ASM count rates for the Suzaku observation period are estimated from the ASM daily count rates 
and summarized in table~1. 
The red and black points in figure \ref{alllc}a (top) show 
the Suzaku data used in this paper, with 0.5--10 keV fluxes measured with the
XIS (absorption is not corrected), $F_{0.5-10}$, and the 15--20 keV count rates in PIN, respectively. 
The hardness ratio of the 15--20~keV PIN count rate 
to the 0.5--10 keV fluxes ($\times 10^{9}$) 
are plotted in red points in figure \ref{alllc}a (bottom). 
The heterogeneous hardness ratio is used to 
compensate for the differences in count rates of XIS for different observation modes.  
Both the ASM and Suzaku data show an increase 
in the flux and a decrease in the hardness ratio in the middle of 2009.

Cyg X-1 is considered to be in the high/soft state 
when the ASM hardness ratio is $\lesssim$ 0.8, 
according to the criterion Fender et al. (2006) proposed. 
Since all the present 25 observations have the hardness ratio $>$ 0.8, 
none of them can be classified as the high/soft state. 
The moderate softening observed in 2009 is 
regarded as a ``failed transition'', 
wherein the source evolved from the low/hard state 
possibly to the hard intermediate state. 

For a more detailed state identification, 
the 25 observations are arranged in a hardness-luminosity diagram in figure \ref{alllc}b. 
Following Dunn et al. (2010), 
we used the luminosity in 3--10 keV, $L_{3-10~\rm{keV}}$, 
divided by the Eddington luminosity, $L_{\rm{Edd}} = 1.5 \times 10^{38}$$( 14.8 M_{\odot} / M_{\odot})$ erg s$^{-1}$ given 
the hydrogen fraction of 0.7, and the hardness between 6--10 keV and 3--6 keV, $L_{6-10~\rm{keV}}/L_{3-6~\rm{keV}}$, 
where $L_{6-10~\rm{keV}}$ is the luminosity in 6--10 keV. 
%
Our data mostly sample the top of the low/hard state ``stalk'' on the hardness-intensity
diagram, where the hardness stays approximately constant 
while the intensity increases. 
In figure~2b, 
several of the 25 data points, 
corresponding to the failed transition mentioned above, 
are seen to connect into the hard intermediate state,
where the hardness decreases markedly for little increase in intensity
(states as defined in Homan \& Belloni 2005). 
Spectra at the top of the stalk (Obs. 1 and 25) 
can either be classified as bright low/hard
state or as hard intermediate state. We classify them as hard
intermediate state in the rest of this paper. 

\subsection{The XIS data}

\subsubsection{Processing of the XIS data} 

The operation modes of the XIS are summarized in table \ref{xismode}.
Among the four XIS cameras, XIS2 was operational until Obs.~2., 
while it stopped working since 2006 November.   
To reduce piled-up events, 
the 1/8 or 1/4 window mode was employed, together with burst options (Koyama et al. 2007). 
One of the XIS data in Obs. 5, 9, 24, and 25 are taken in a timing mode, or Parallel(P)-sum
mode, which is one of the clocking modes in the XIS. 

We begin with the standard screening criteria of the XIS as follows:
(1) the XIS Grade \citep{Koyama2007} should be 0, 2, 3, 4, or 6, (2)
the time interval after an exit from the South Atlantic Anomaly should
be longer than 436 s, and (3) the object should be at least
5$^{\circ}$ and 20$^{\circ}$ above the dark and sunlit Earth rim,
respectively.  
Data taken with the time mode were processed with a different Grade selection\footnote{see http://www.astro.isas.ac.jp/suzaku/analysis/xis/ \\psum\_recipe/Psum-recipe-20100724.pdf}: Grade 0 (single
event), as well as 1 and 2 (double events), were used.  These data
have a time resolution of 7.8 ms owing to onboard summation over one
of the two dimensions.

Since Cyg X-1 is very bright and the combination of the XIS and the
XRT provides a relatively broad spatial resolution of $\sim2'$ in a
half-power diameter (Serlemitsos et al. 2007), it is difficult to
extract non X-ray background (NXB) from the outskirt of the XIS image.
Therefore, we estimated the NXB level using 
the events acquired in an north-ecliptic pole
region, dominated by the NXB (\cite{Kubota2010}). 
The NXB was then found to contribute
$\lesssim 0.1$\% to the Cyg X-1 data even at the highest energy end,
$\sim$ 10 keV.  Therefore, in the present XIS analysis, the NXB
subtraction is not needed.

The source brightness instead demanded additional two screening steps:
to exclude those periods when the telemetry of the XIS was saturated,
and to avoid piled-up photons of the XIS.  
The former was discarded by
using the house keeping data, {\tt ae*xi*\_tel\_uf.gti}.  The latter
was practically realized by excluding a center region of the image, to
minimize pileup effects at the sacrifice of photon statistics.  This
``core excision'' method was calibrated for the XIS data analysis in
data-oriented way (Yamada et al. 2012).  We also corrected the XIS
image for smearing effects due to thermal wobbling of the spacecraft
\citep{Uchiyama2008}. 


\begin{table}[htbp]
 \caption{Window/burst options applied to the XIS.}
 \label{xismode}
 \begin{center}
  \begin{tabular}{ccccccc}
   \hline\hline
  & \multicolumn{1}{c}{XIS0} & \multicolumn{1}{c}{XIS1}  & \multicolumn{1}{c}{XIS3}  \\[1mm] 
$N$$^*$          & Mode & Mode &  Mode \\[1mm]                      
\hline	
1$^\dagger$ & 1/8W 1.0 s & 1/8W 1.0 s & 1/8W 1.0 s \\
2$^\dagger$ & 1/8W 0.3 s & 1/8W 0.3 s & 1/8W 0.3 s \\
5 & timing & std$^\ddagger$ & std \\
9 & timing & std & timing \\
24 & std & std & timing \\
25 & std & std & timing \\                         
\hline\hline
\end{tabular}
\end{center}
\begin{itemize}
\setlength{\parskip}{0cm} %
\setlength{\itemsep}{0cm} %

\item[$^*$] The observation number used in table 1.
\item[$^\dagger$]  The modes and exposures of XIS2 are the same as those of XIS0. 
\item[$^\ddagger$] ``std'' means a standard mode for a bright source: 1/4 window mode and 0.5 s burst option. 
In the other 20 observations, the XIS cameras were all operated in ``std''. 
\end{itemize}
\end{table}

\subsubsection{Lightcurves and spectra of the XIS}
\label{sec:specxis}

Since Cyg~X-1 has a supergiant companion, the accreting gas is
supplied by its strong stellar winds. 
This causes episodes of increased absorption, or ``dips'', which are seen as decreases in soft
X-rays at near the superior conjunction of the black hole (e.g.,
\cite{Kitamoto1984}).  Since we are interested in the intrinsic
emission properties of Cyg X-1, these dipping periods were removed as
detailed in Appendix~\ref{sec:dip}. 

We then extracted events from an annulus with the inner radius 
of $R_{\rm3\%}$ as shown in table \ref{xisinfotbl} in Appendix \ref{sec:att-pl} 
and the outer radius of 4$'$, 
where $R_{\rm x\%}$ refers to the radii at which the pileup fraction is $x$\%.   
These details are described in Appendix \ref{sec:att-pl}. 
Since adopting much smaller pileup fractions (much stronger core excision) would increase 
systematic uncertainties in the XIS and XRT response \citep{2012PL}, 
the pileup fraction of 3\% for spectral analysis is practically reasonable. 

Figure \ref{hilolc} shows 0.5--10 keV light curves of XIS3 extracted
from a circle with a radius of $4'$, 
since the count rate is less susceptible to pileup than spectral shape.  
In Obs.~5, 9, 24, and 25, 
either the XIS0 or XIS3 data acquired 
in the timing mode were used. 

The response matrices and auxiliary response
files were created with {\tt xissimrmfgen} and {\tt xissimarfgen},
respectively \citep{xissim}; the latter properly takes into account
areas used for  spectral extraction, 
and the widow shapes of the XIS sensors.
The systematic differences in normalization among the four (or three)
XIS sensors are less than $\sim$ 3\% 
(suzakumemo 2008-06\footnote{http://www.astro.isas.ac.jp/suzaku/doc/suzakumemo/suzakumemo-2008-06.pdf}).


\subsection{The HXD data}

\subsubsection{Processing of the HXD data}

The PIN and GSO events were screened by their standard criteria:
the target elevation angle $\geq$ $5^{\circ}$, cutoff rigidity $\geq$ 6~GV, and
500 s after and 180 s before the South Atlantic Anomaly.  The
telemetry-saturated and FIFO-full periods (Kokubun et al. 2007) were
discarded.  In order to utilize the latest GSO calibration results
(Yamada et al.~2011), we have reprocessed all the GSO data with {\tt
hxdpi} implemented in HEAsoft 6.9 or later, together with the
corresponding gain history parameter table, {\tt ae*gpt*fits}.

The NXB of the HXD was subtracted from the raw data 
as described in Fukazawa et al. (2009). 
The cosmic X-ray background was not subtracted 
from either the PIN or GSO data, since its contribution
integrated over the HXD filed of view is $<$ 0.1\% of the signal in
any energy range.  The basic information on the HXD data is almost the
same as that in Paper~II, so that the details are summarized in
Appendix~\ref{sec:hxdcr}.

\subsubsection{Lightcurves and spectra of the HXD data}
\label{sec:spechxd}

We used the responses of PIN and GSO provided by the detector team.
The NXB of PIN was generally less than 1\% of the source signals
therein, and at most $\sim$ 10 \% at $\sim$ 60 keV.  The NXB of GSO
became comparable to the source signals at $\sim$ 150 keV.  Since
systematic uncertainties in the NXB of GSO are at most $\sim$ 3\%
(Fukazawa et al. 2009), we use the spectra up to $\sim$ 300 keV.  More
detailed studies of the spectra and timing properties of the HXD data
were reported in Paper~II.

\begin{figure}[htbp]
    \begin{center}
      \includegraphics[width=0.49\textwidth]{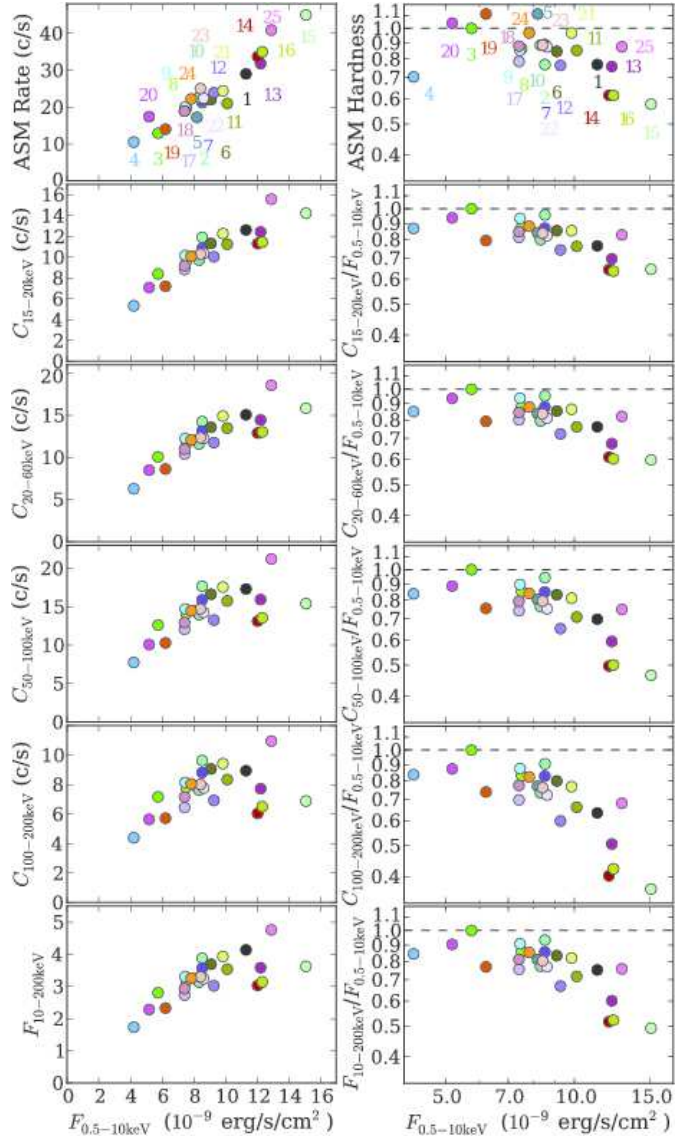}
            \end{center}
    \caption{(left; top to bottom) The 1.5--12 keV ASM count rate, the 15--20 keV PIN count rate, that in 20--60 keV, 
    the 50--100 keV GSO count rate, that in  100--200 keV, and the 10--200 keV signal flux, all plotted against the 0.5--10 keV signal flux, $F_{0.5-10}$. The 25 observations are specified by colors.  
    (right) Hardness ratios in various energy bands, all plotted against $F_{0.5-10}$. The top panel show the 5--12 keV to 1.5--3 keV ratios for the RXTE ASM, 
    while the lower 5 panels give ratios of the count rates in the corresponding left-column panel to $F_{0.5-10}$. All are normalized to the value of Obs. 3. }
    \label{avepara}
\end{figure}

\subsection{Correlations of fluxes in different energy bands}


The left side panel in figure \ref{avepara} presents the relation of 
the flux in 0.5--10 keV, $F_{0.5-10}$, to the count rates of the ASM, 
PIN and GSO, as well as the 10--200 keV flux. 
All of them show a positive correlation with
$F_{0.5-10}$, which means that the overall source behavior, to the
first approximation, is controlled by a single parameter; i.e., the
spectrum normalization, which is considered to reflect the mass accretion rate. 
However, 
the spectral shape is also varying to some extent, 
since the 0.5--10~keV flux changes more than the 100--200~keV flux. 

To better visualize these spectral changes, we show hardness ratios in the
right hand panel of figure
\ref{avepara}.  When the 0.5--10 keV flux exceeds $\sim 7 \times
10^{-9}$ erg s$^{-1}$ cm$^{-2}$ (marking the top of the low/hard state
``stalk'' in the hardness-intensity diagram in figure \ref{alllc}b) the
hardness ratio to all bands starts to decrease, and this effect
increases towards higher energies, suggesting that 
not only a spectral slope steepens but also a high energy cutoff becomes lower.

\subsection{Estimating the Eddington ratios} 

To assess the Eddington ratios for our sample. 
we employed the same double Comptonization model 
as used in Makishima et al.~2008 with the neutral column density fixed at $7 \times 10^{21}$cm$^2$. 
Since the model reproduced the wide-band spectra within several percents, 
we evaluated absorbed and unabsorbed 0.1--500 keV fluxes
by extrapolating the best-fit model spectra. 
Considering isotropic emission, 
we obtained absorbed and unabsorbed 0.1--500 keV luminosities, 
$L_{0.1-500}$, by multiplying obtained fluxes by 4$\pi$$D^2$, 
and listed them in table~1.  
Our data sample a range of 0.8--2.8\% of $L_{\rm{Edd}}$, 
which are grossly typical values observed from black hole binaries in the low/hard state. 

\section{Long-term Spectral Analysis}

\subsection{Wide-Band Spectra}
\label{sec:wbs}

Removal of instrumental responses (conventionally ``unfolding'') is
useful to make spectral changes more directly visible, without being
affected by the instrumental effects such as energy-dependent detector
efficiencies.  This is usually performed after finding the best-fit
models, because results of deconvolution are known to depend on the
employed model (so-called obliging effect).  However, this effect is 
considered negligible
even if we use next-to-best fitting models,
as long as the model and the data are both free from features
which are comparable to or narrower than the instrumental energy
resolution. 
Therefore, we try to find approximately good fitting models,
and perform the deconvolution. 

In fitting the broad-band spectra,  
we used the response and auxiliary files described in section 
\ref{sec:specxis} for the XIS and \ref{sec:spechxd} for the HXD.
The cross normalization between the XIS and HXD was fixed at the standard 
value of 1.17, of which the uncertainty is
estimated to be less than $\sim$ 5\% based on the Crab Nebula data
(suzakumemo 2008-06).  
Then, we fitted separately the XIS and HXD spectra with conventional models,
\texttt{diskbb + powerlaw} and \texttt{compps}, respectively, where
{\tt diskbb} is a multicolor disk model (Mitsuda et al. 1984;
Makishima et al. 1986) and {\tt compps} is one of the Comptonization
models valid for $\tau < 3$ \citep{1996ApJ...470..249P}. 
Using these models, we conducted the spectral deconvolution. 
The deconvolved spectra are changed by less than $<$1\% when these different
models are used.

\begin{figure*}[htbp]
    \begin{center}
      \includegraphics[width=0.98\textwidth]{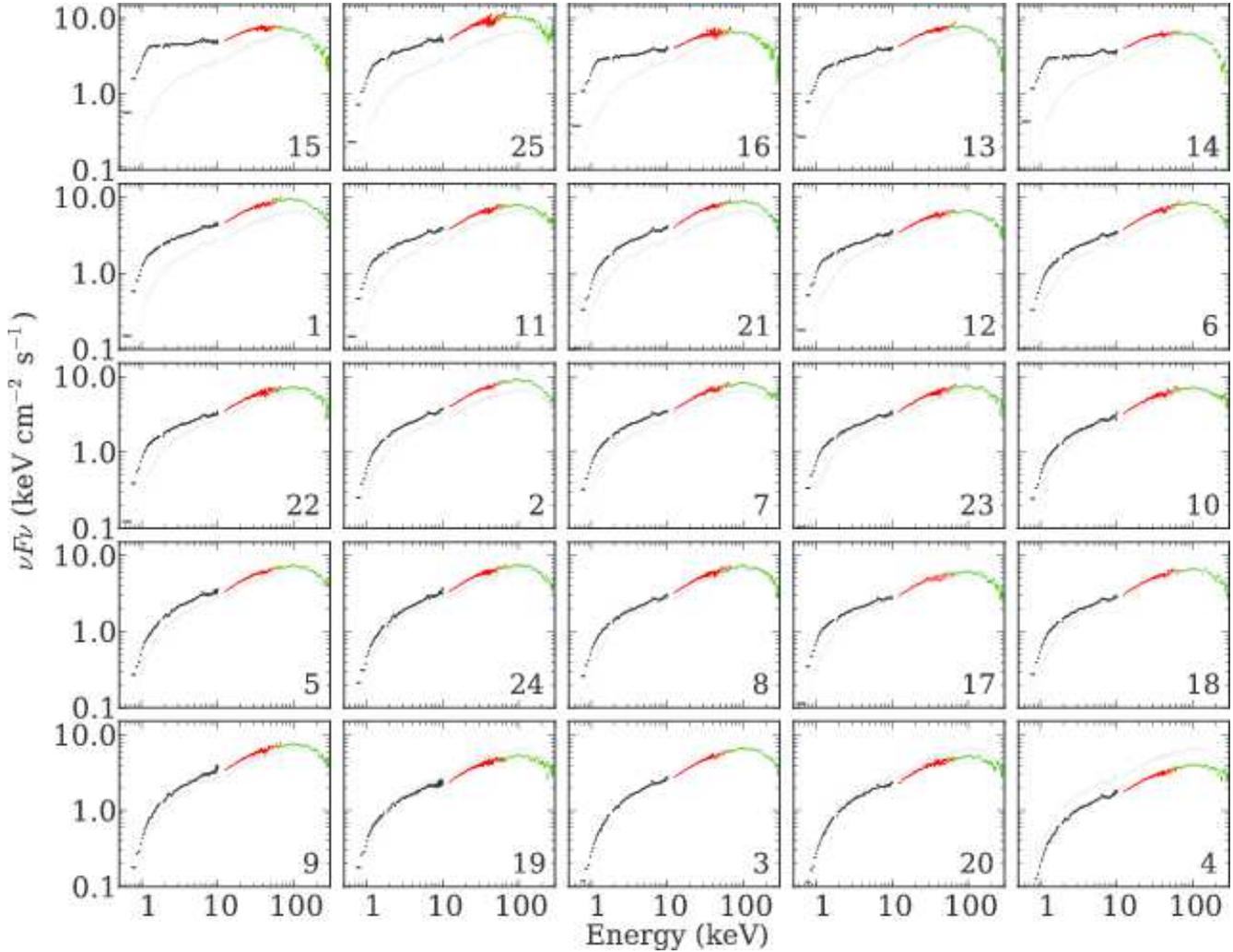}
            \end{center}
    \caption{Background-subtracted Suzaku spectra of Cyg X-1 from the 25 observations, 
    shown in the $\nu F_\nu$ form after removing the detector responses. 
    The spectra of XIS, PIN, and GSO are shown in black, red, and green, respectively. 
    The XIS data refer to XIS3, except in Obs.~9 in which XIS1 in used, 
    and Obs.~24 and 25 wherein XIS0 is employed. The spectra of Obs.~3 are superposed in light gray as a reference. 
    } 
    \label{alleeuf}
\end{figure*}

Figure \ref{alleeuf} shows the deconvolved 0.5--300 keV spectra 
of the 25 Suzaku observations of Cyg X-1, 
presented in the $\nu F \nu$ form,
and sorted in terms of decreasing soft X-ray flux
($F_{0.5-10}$ given in table 1). 
We overlay the 
hardest spectrum (Obs.~3) in grey on all the data. 
This visualizes systematic changes 
in the spectral shape 
as a function of $F_{0.5-10}$, 
which are suggested by the hardness ratios of figure \ref{avepara}. 
We also note that the four spectra on the top row in figure~5, 
with the highest $F_{0.5-10}$ flux, correspond to the hard
intermediate state spectra of figure~\ref{alllc}b, 
while the rest are in the low/hard state.

At a first glance, 
these $\nu F \nu$ spectra all show very similar 
shapes, monotonically rising up to $\sim$ 100 keV, then sharply falling. 
This is especially true for all the low/hard state spectra. 
These results agree, at least qualitatively, 
with the previous Suzaku studies of Cyg X-1 
conducted in the low/hard and hard-intermediate state 
(Obs. 1: Makishima et al.~2008; Obs. 1-4: Nowak et al.~2011; 
Obs. 7: Fabian et al.~2012). 
 
By inspecting figure \ref{alleeuf} more closely, we can deduce the
following characteristics of the spectral changes which occurred among
the observations on a time scale of days to months.  
\begin{enumerate}
\setlength{\parskip}{0.05cm} %
\setlength{\itemsep}{0.05cm} %
\setlength{\leftskip}{-0.5cm}
\item As the flux increases, the emission below $\sim$ 10 keV becomes more prominent. 
\item At the same time as 1, the 10--100 keV spectral slope steepens and the cutoff
feature at $\sim$ 100 keV moves to lower energies.  
\item Contrary to 2, in the dimmest two observations, Obs.~20 and 4, 
the spectrum softens even though the flux decreases. 
\end{enumerate}
The last property is characteristic of dimmer low/hard state spectra from other  black hole binaries
(e.g. Sobolewska et al. 2011), but this is the first time it has been
identified in Cyg X-1. 
The transition of ``softer when brighter'' into ``harder when brighter'' behavior 
in Cyg X-1 appeared at $L_{\rm{Edd}} \sim 0.01$, 
which agrees with the results of GRO~J1655-40 and GX 339-4 studied by Sobolewska et al. 2011. 
Thus, it implies that some changes in accretion at $L_{\rm{Edd}} \sim 0.01$ 
might be a universal property of black hole binaries. 

\begin{figure*}[htbp]
    \begin{center}
      \includegraphics[width=0.98\textwidth]{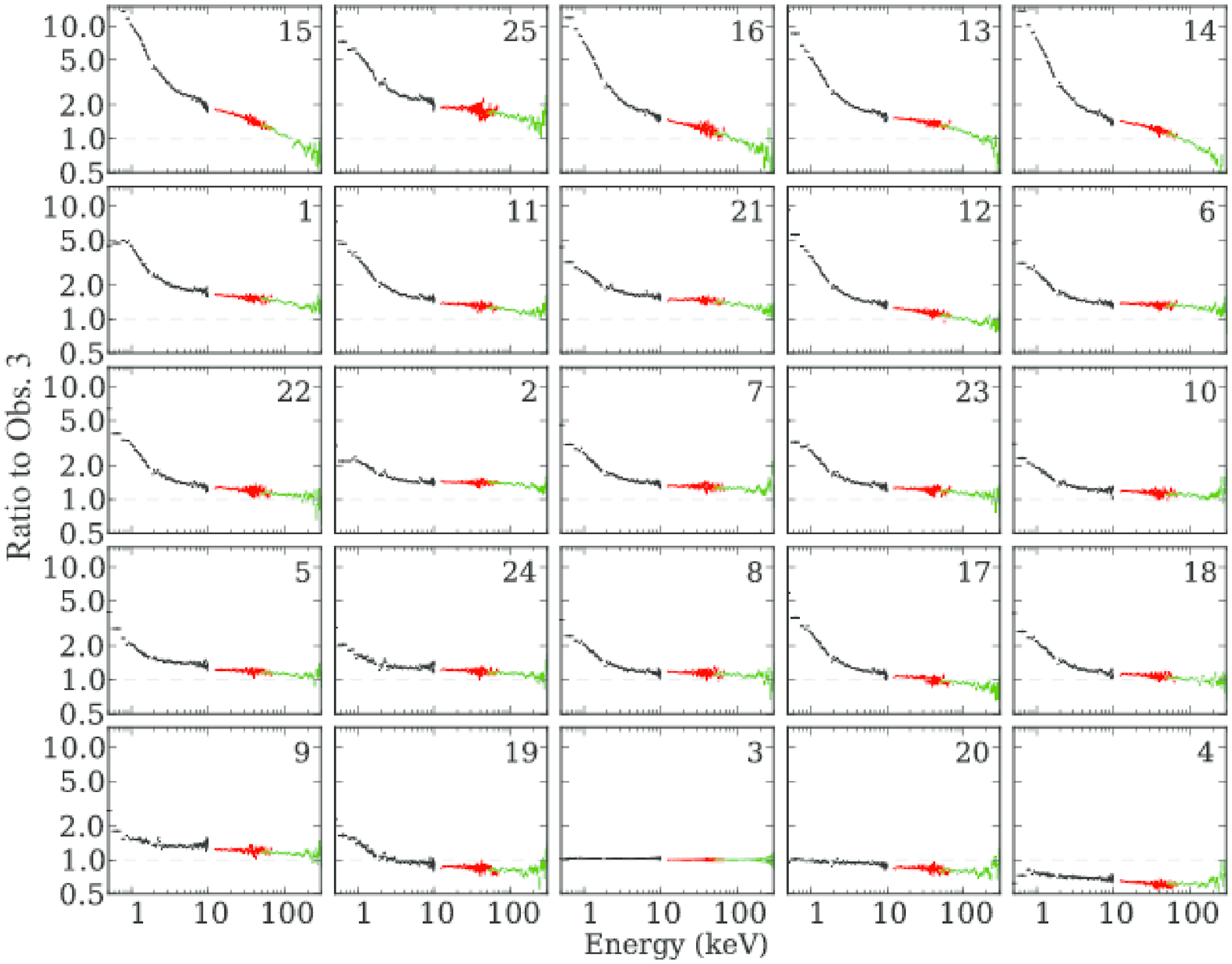} 
            \end{center}
    \caption{The same wide-band spectra as shown in figure \ref{alleeuf}, but all normalized to that of Obs.~3.}
    \label{eeufratio}
\end{figure*}

\subsection{Spectral Ratios on a Time Scale of $>$ Days}

To more directly examine the spectral differences among the
observations, we created spectral ratios in figure \ref{eeufratio},
again using the hardest spectrum (Obs. 3) as a reference.  The
properties (1)--(3) noticed in subsection \ref{sec:wbs} are more
clearly visualized.  The ratios are larger than 1.0 for all data below 10~keV
(apart from the two lowest luminosity spectra, Obs. 20 and 4), but is
always larger at 1~keV than at 10~keV. Thus the long term variability
is largest at soft energies.

This long term variability clearly has (at least) two components.  For
example, in Obs.~15 (the brightest hard intermediate state), 
the ratio shows a sharp upturn below 2~keV, 
then a steady decline in the 3--30~keV bandpass and 
a sharper decline above 100~keV. 
To zeroth order this is what is expected when
the seed photons illuminating the Compton region increase by more than
the power dissipated in the hot electrons (Zdziarski et al. 2002; Zdziarski \& Gierli\'{n}ski 
2005) as the photon index steepens, and electron
temperature drops.

A further inspection of the $\nu F_\nu$ spectrum of Obs.~15 in figure~\ref{alleeuf} 
shows that the spectral changes above 3~keV 
cannot be explained by considering the two components (disk and Comptonization) only. 
In brighter spectra, 
the ratio starts rising abruptly at 3--4 keV 
toward lower energies, where the disk emission should still be negligible.  
Therefore, the data indicate the presence of an additional soft
continuum component with a spectral index of $\Gamma\sim 2$ which
dominates below 10~keV.


\section{Short-term Spectral Analysis} 
\label{sec:is}

We first explore the rising component below 10~keV. 
If this contains the disk, 
then it can have very different variability properties from
the corona.  
Hence we look at the rapid (1--2s) variability in the XIS,
and use this to separate out the corona from the slower variability of
the disk component. 

\subsection{Definition and preparation} 
\label{sec:is_def}

We use ``intensity-sorted spectroscopy'', as defined in Paper~I to
study the spectral variability on a time scale of 1--2 s. 
Specifically, we define high-flux phase as 
\begin{equation}
\{  t  \mid  C(t)  >  ( 1 + f )  ~ \overline{C(t)}_{T}  \},
\end{equation}
while low-flux phase as 
\begin{equation}
\{  t  \mid  C(t)  \leqq  ( 1 - f ) ~ \overline{C(t)}_{T}  \}, 
\end{equation}
where $C$($t$) and $t$ are the count rate and the event arriving time, respectively, 
$f$ is a threshold to determine the high and low intensities, 
$T$ is an interval over which $C$($t$) is averaged, 
and $\overline{C(t)}_T$ denotes the average count rate over the interval of $t - T/2 <  t  < t + T/2$. 
The gap between the high and low phases is defined as an intermediate-flux phase.  
This differs slightly from Paper~I in that we calculate
 $\overline{C(t)}_T$ as a running average at every $t$, 
while in Paper~I it was set every 200 seconds in a discrete manner. 

When conducting the intensity-sorted spectroscopy with a single detector, 
we need to set $f$ rather high to avoid Poisson noise; 
otherwise, the high and low spectra would pick up the effects of statistical fluctuations. 
However, in the present case, this problem can be avoided 
by judging the high and low phases using a detector, and 
accumulating the high and low spectra using the other detectors. 
Specifically, 
we make the lightcurve $C(t)$ from one of the XIS cameras, 
and determine the high/low flux phase based on equation~(1) and (2). 
Then, according to the criteria, 
the high-phase and low-phase events of the other XIS cameras are obtained. 

For creating $C(t)$, the 0.5--10 keV count rate of XIS3 was used after
excluding the dipping periods.  We set $f$ at $0.05$ as a compromise
between photon statistics and the contrast in intensity, and $T$ at 64
s.  The color coding on figure \ref{hilolc} gives the high (dark
grey), intermediate (light grey), and low (black) phases determined
for all observations according to these parameters.

\subsection{Spectral changes on a time scale of 1--2 s}
\label{sec:defhl}
    
\begin{figure*}[htbp]
    \begin{center}
      \includegraphics[width=0.98\textwidth]{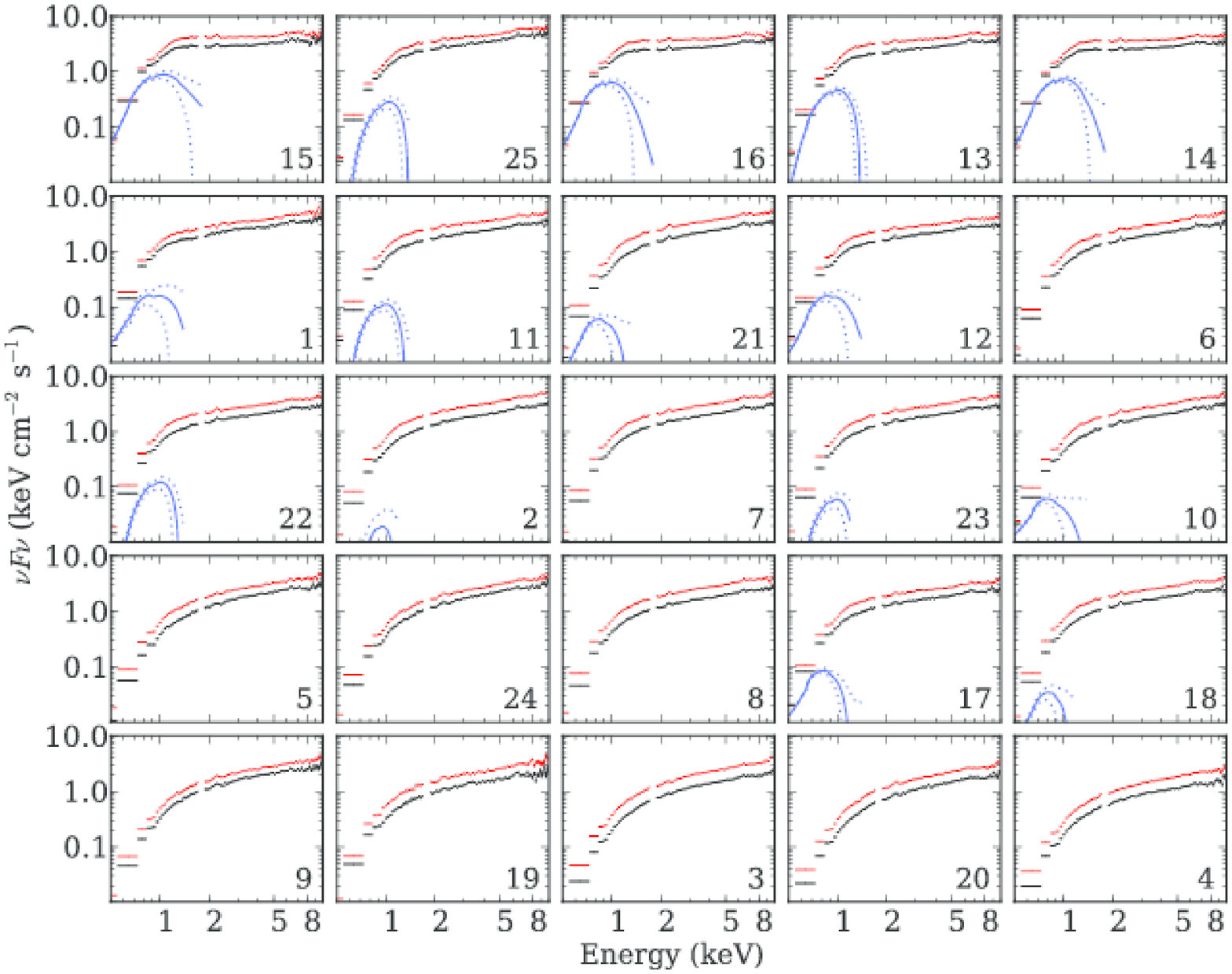}
            \end{center}
    \caption{The intensity-sorted $\nu F_\nu$ spectra of Cyg X-1, taken with XIS1. 
    The high and low phase spectra are shown in red and black, respectively. 
    The estimated disk component is shown in blue, with their 90\% errors in dotted lines.} 
    \label{hiloeeuf}
\end{figure*}
 
We first looked in detail at the low energy data.  We sorted the XIS1
events into the high and low phases according to the criteria in
section \ref{sec:is_def}, and obtained the results shown in figure
\ref{hiloeeuf} in $\nu F_\nu$ form.  To zeroth order, the low and high
spectra exhibit similar shapes; 
i.e., the source variation on the 1--2 s
time scale occurs primarily keeping the spectral shapes constant.

\begin{figure*}[htbp]
    \begin{center}
      \includegraphics[width=0.98\textwidth]{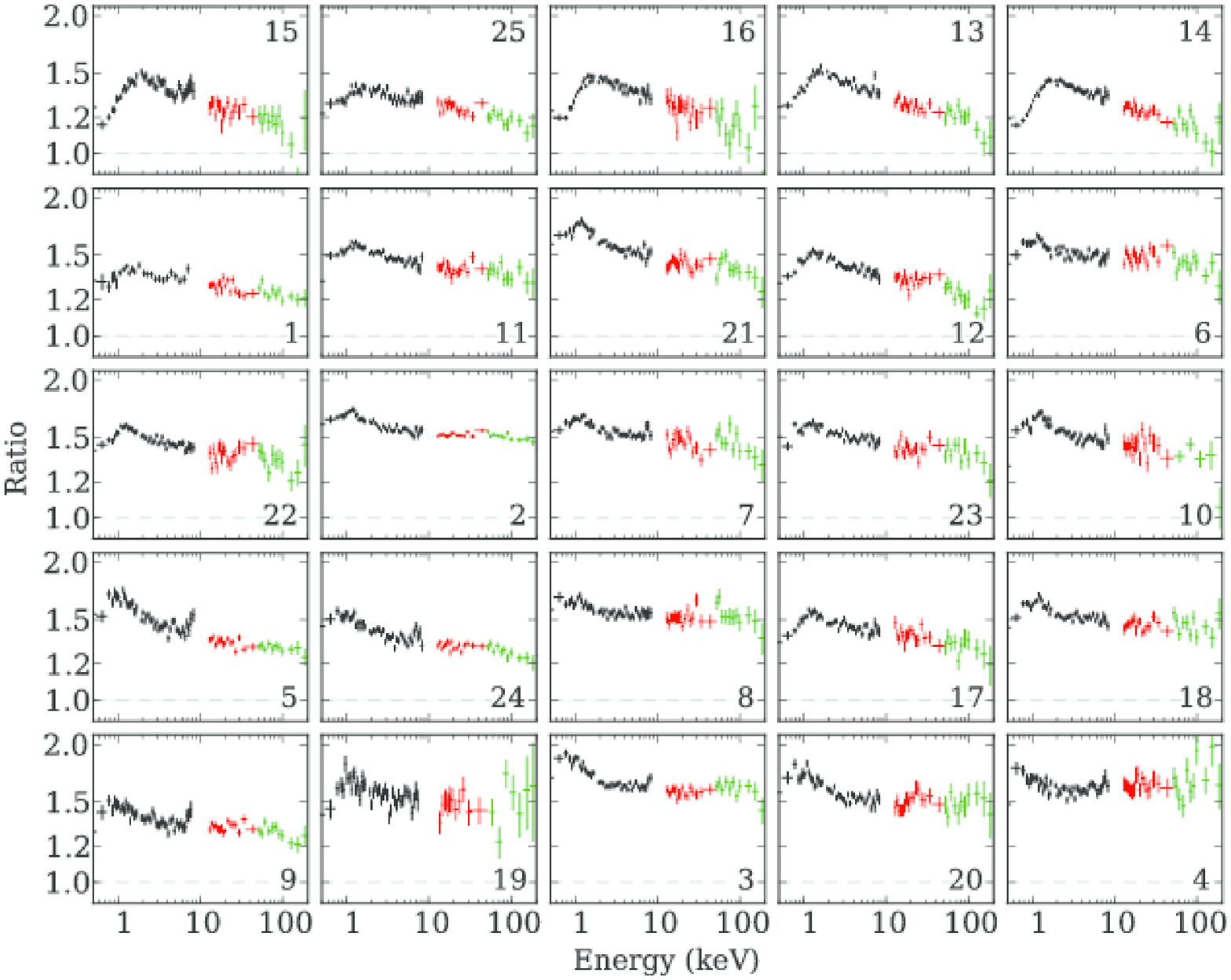}      
            \end{center}
    \caption{The ratios between a pair of low-flux and high-flux spectra of individual observations 
    on a time scale of 1--2 s. The colors are the same as used in figure \ref{alleeuf}. 
    } 
    \label{hiloratio}
\end{figure*}

We then extend the energy band using the PIN and GSO data. We form the
high and low spectra for each observation across the entire energy
band and show the ratio of the high-phase spectrum to the low-phase
one in figure \ref{hiloratio}.  These ratio plots reveals more subtle
variations. There is a clear peak in the variability at energies
around $\sim 0.5$--$1$~keV in all spectra. 
The hard intermediate state and bright low/hard state observations (top rows) 
show a sharp dip in variability at energies below the peak, 
and a moderate decrease to higher energies. 
By contrast, 
the very lowest luminosity spectral ratios (bottom row) look like
the typical low/hard state but with the peak shifted to lower
energies, and the variability drop to high energies appears to stop, 
settling to a constant variability over 3--300~keV. 

\subsection{Quantification of the constant disk emission}
\label{sec:quanhl}

We now quantify the short-term variability suppression in $E
\lesssim 2$ keV found in some observations.  
Assuming that the high-phase spectrum, $H(E)$, and
the low-phase spectrum, $L(E)$, are decomposed into a sum of a stable
soft component $d(E)$ and a varying harder component, to be denoted
$h(E)$ in the former and $l(E)$ in the latter, then this can be
expressed as
\begin{equation}
H(E) = w(E) \left( d(E) + h (E) \right)
\end{equation}
\begin{equation}
L(E) = w(E) \left( d(E) + l (E) \right)
\end{equation}
where $w(E)$ is photoelectric absorption.  
The spectral ratio $R(E)$ is then written as
\begin{equation}\label{eq:hl}
R(E) \equiv \frac{H(E)}{L(E)} = \frac{ d(E) + h (E) }{ d(E) + l (E)}.
\end{equation}
Since we have three unknown quantities, $d(E)$, $h(E)$, and $l(E)$, 
it is impossible to uniquely solve equation (\ref{eq:hl}) for them.  
Therefore, let us assume that 
the ratio at $E > 2$ keV can be approximated, 
using two parameters $\alpha$ and $\beta$, as  
\begin{equation}
R(E) \simeq \frac{h(E)}{l(E)} = \alpha E^{\beta}.
\end{equation}
Supposing further that this relation between $l(E)$ and $h(E)$ can be extrapolated down to $\sim$ 0.5 keV, 
equation (3) and (4) can be solved for $d(E)$ as 
\begin{equation}\label{eq:disk}
w(E) d(E) = \frac{ \alpha E^{\beta} L(E) - H(E) }{ \alpha E^{\beta} -1 }. 
\end{equation}

\begin{table}[htbp]
\caption{Results of fitting the low-to-high spectral ratio in figure~\ref{hiloratio} with a single powerlaw.}
\label{fithiloratio}
\begin{center}
\begin{tabular}{ccccccc}
\hline\hline
N &   $\alpha$$^*$ & $\beta$$^*$ & $\chi^2$(d.o.f)  \\
\hline	
1 & 1.429$\pm$0.027 & -0.060$\pm$0.020 & 41.8(33) \\
2 & 1.693$\pm$0.014 & -0.068$\pm$0.009 & 30.4(33) \\
3 & 1.749$\pm$0.018 & -0.065$\pm$0.011 & 41.0(33) \\
4 & 1.596$\pm$0.023 & -0.011$\pm$0.015 & 34.1(33) \\
5 & 1.650$\pm$0.024 & -0.106$\pm$0.016 & 37.4(33) \\
6 & 1.556$\pm$0.018 & -0.028$\pm$0.013 & 41.2(33) \\
7 & 1.622$\pm$0.016 & -0.054$\pm$0.011 & 53.8(33) \\
8 & 1.601$\pm$0.019 & -0.034$\pm$0.013 & 20.7(33) \\
9 & 1.445$\pm$0.024 & -0.057$\pm$0.017 & 35.5(33) \\
10 & 1.663$\pm$0.020 & -0.086$\pm$0.013 & 43.2(33) \\
11 & 1.596$\pm$0.016 & -0.062$\pm$0.011 & 39.9(33) \\
12 & 1.567$\pm$0.016 & -0.086$\pm$0.011 & 26.8(33) \\
13 & 1.582$\pm$0.015 & -0.088$\pm$0.010 & 34.8(33) \\
14 & 1.493$\pm$0.013 & -0.060$\pm$0.010 & 19.6(33) \\
15 & 1.484$\pm$0.020 & -0.035$\pm$0.015 & 33.9(33) \\
16 & 1.508$\pm$0.015 & -0.058$\pm$0.012 & 30.2(33) \\
17 & 1.543$\pm$0.017 & -0.052$\pm$0.012 & 58.0(33) \\
18 & 1.592$\pm$0.016 & -0.043$\pm$0.011 & 39.5(33) \\
19 & 1.735$\pm$0.041 & -0.121$\pm$0.026 & 38.6(33) \\
20 & 1.734$\pm$0.022 & -0.111$\pm$0.014 & 49.3(33) \\
21 & 1.736$\pm$0.018 & -0.098$\pm$0.011 & 29.7(33) \\
22 & 1.587$\pm$0.015 & -0.070$\pm$0.011 & 40.7(33) \\
23 & 1.615$\pm$0.016 & -0.062$\pm$0.011 & 43.5(33) \\
24 & 1.519$\pm$0.019 & -0.090$\pm$0.014 & 29.8(33) \\
25 & 1.426$\pm$0.018 & -0.053$\pm$0.014 & 30.7(33) \\
\hline\hline
\end{tabular}
\end{center}
\begin{itemize}
\setlength{\parskip}{0cm} %
\setlength{\itemsep}{0cm} %
\item[$*$] The values of $\alpha$ and $\beta$ are defined as $\alpha E^{\beta}$. 
\end{itemize}
\end{table}

\begin{table}[htbp]
\caption{Derived parameters of the stable disk emission.}
\label{fitdiskpara}
\begin{center}
\begin{tabular}{llllr}
\hline\hline
N & $R^{\rm{tr}}_{\rm{in}}$ ($R_{\rm{g}}$) & $T^{\rm{tr}}_{\rm{in}}$ (keV) & $L^{\rm{tr}}$$^*$ & $\chi^2$(d.o.f) \\[2mm]                      
\hline 
1 &        17.5$_{-8.6}^{+59.1}$ & 0.13$_{-0.04}^{+0.03}$ & 5.9$_{-2.2}^{+19.3}$ & 0.01(11)\\ 
10 &     39.6$_{-29.8}^{+134.6}$ & 0.10$_{-0.03}^{+0.04}$ & 9.1$_{-6.9}^{+44.5}$ & 0.07(11)\\
11 &     12.5$_{-4.0}^{+6.3}$ & 0.13$_{-0.02}^{+0.02}$ & 3.2$_{-0.8}^{+1.1}$ & 11.48(11) \\
12 &     24.8$_{-12.6}^{+49.7}$ & 0.12$_{-0.03}^{+0.03}$ & 8.4$_{-3.8}^{+17.7}$ & 0.31(11)\\
13 &     32.9$_{-8.8}^{+10.8}$ & 0.13$_{-0.01}^{+0.01}$ & 17.8$_{-4.0}^{+4.8}$ & 17.92(15) \\
14 &     26.3$_{-8.8}^{+17.1}$ & 0.14$_{-0.02}^{+0.02}$ & 18.5$_{-5.4}^{+10.7}$ & 1.70(17) \\
15 &     17.0$_{-5.9}^{+16.8}$ & 0.16$_{-0.03}^{+0.02}$ & 13.4$_{-3.5}^{+10.7}$ &  6.34(17) \\
16 &     30.3$_{-12.9}^{+25.4}$ & 0.14$_{-0.02}^{+0.02}$ & 20.0$_{-7.6}^{+15.6}$ & 1.86(17) \\
17 &     76.0$_{-43.2}^{+69.8}$ & 0.09$_{-0.01}^{+0.02}$ & 22.7$_{-13.8}^{+26.1}$ & 4.08(10) \\
18 &     60.5$_{-45.2}^{+186.8}$ & 0.08$_{-0.02}^{+0.03}$ & 12.0$_{-9.2}^{+55.3}$ & 0.47(~6) \\
21 &     41.1$_{-30.1}^{+73.7}$ & 0.10$_{-0.02}^{+0.04}$ & 9.4$_{-6.9}^{+21.0}$ & 1.12(10) \\
22 &     13.1$_{-4.6}^{+7.9}$ & 0.13$_{-0.02}^{+0.02}$ & 3.5$_{-0.9}^{+1.5}$ & 9.73(11) \\
25 &     9.9$_{-2.5}^{+5.7}$ & 0.16$_{-0.02}^{+0.02}$ & 4.1$_{-0.5}^{+1.3}$ &  9.67(11) \\ [2mm]
M1$^\dagger$ & 7.4$_{-1.8}^{+6.2}$ & 0.19$_{-0.02}^{+0.01}$ & 4.6$_{-2.0}^{+11.1}$ &  \\ [2mm]
\hline\hline
\end{tabular}
\end{center}
\begin{itemize}
\setlength{\parskip}{0cm} %
\setlength{\itemsep}{0cm} %
\item[$^*$] Disk luminosity in $10^{36}$ erg s$^{-1}$. 
\item[$^\dagger$] Values derived by spectral fitting in Obs.~1 in Makishima et al. (2008). 
\end{itemize}
\end{table}

To determine $\alpha$ and $\beta$, we fit the spectral ratios in figure \ref{hiloratio} with a single powerlaw
over the 2--4 keV range, where the continua are not affected by the Fe-K line and edges, 
and summarize the best-fit parameters in table \ref{fithiloratio}. 
Then, using equation (\ref{eq:disk}) and the obtained values of $\alpha$ and
$\beta$, we calculated $w(E)d(E)$ and show the results in figure
\ref{hiloeeuf} (blue).  Thus, in at least 10 out of the 25 data sets
(almost all the hard intermediate state and bright low/hard state and
almost none of the dim low/hard state), we significantly detected a
soft and stable spectral component that is responsible for the
variability suppression in $\lesssim$ 2 keV.  The derived component
exhibits a shape which is similar to that expected from
low-temperature disk emission (Paper~I). 

\begin{figure}[htbp]
    \begin{center}
      \includegraphics[width=0.45\textwidth]{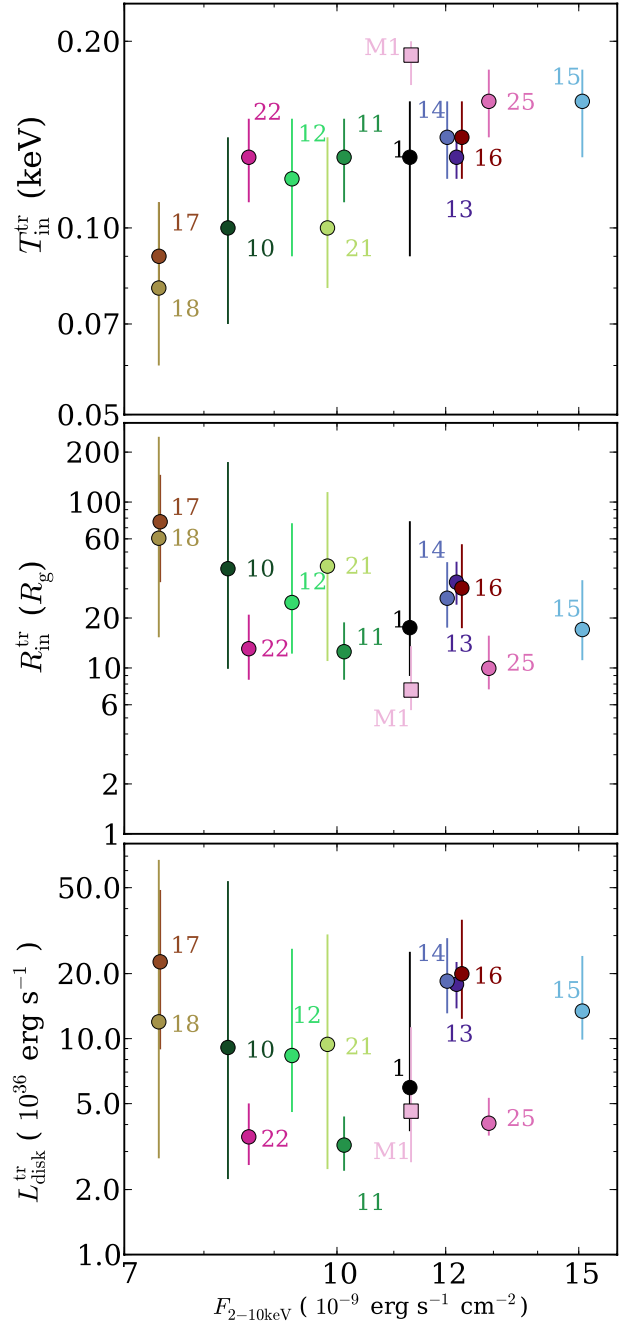}
            \end{center}
    \caption{The relation between the 0.5--10 keV flux and (a) the disk temperature, 
    (b) the inner disk radius, and (c) the disk luminosity, listed in table \ref{fitdiskpara}. The result of Paper~I for Obs.~1 is shown by pink squares.} 
    \label{diskplot}
\end{figure}

To parameterize the shape of this constant component, $w(E)d(E)$, we
fitted them with a {\tt wabs * diskbb} model, with the neutral column
density in $w(E)$ fixed at $7 \times 10^{21}$cm$^2$.  The errors of
$w(E)d(E)$ were defined as the two extreme values specified by
allowable ranges of $\alpha$ and $\beta$, and are indicated in dotted
lines in figure 7.  The free parameters are the inner radius
$R_{\rm{in}}$ and the inner disk temperature $kT_{\rm{in}}$, which
specify the total disk luminosity $L_{\rm{disk}} = 4 \sigma \pi
R_{\rm{in}}^2 T_{\rm{in}}^4$, where $\sigma$ is the Stefan-Boltzmann
constant.  As summarized in table \ref{fitdiskpara} and figure
\ref{diskplot}, the fit was generally successful, and yielded
$kT_{\rm{in}} \sim 0.1$--$0.2$ keV, and $R_{\rm{in}}
\sim100$--$1000$~km, where $k$ is the Boltzmann constant. 
To distinguish these from those usually obtained
by fitting entire spectra, we put a subscript of ``tr'', referring to a
transmitted disk emission, in table \ref{fitdiskpara} and figure \ref{diskplot}. 

We note that correcting the {\tt diskbb} radii 
for a stress-free inner boundary condition and 
color temperature correction 
(assuming 0.41 and 1.7, respectively: Kubota et al.~2001) 
increases the radii only slightly by a factor of 1.18. 
However, the stress free inner boundary condition explicitly 
assumes that the disk extends down to the last stable orbit. 
A truncated disc may well have stress on its innermost orbit, 
making it more like the {\tt diskbb} assumptions 
(Li et al.~2005; Gierli\'{n}ski, Done \& Page~2008).
Assuming that there is no stress correction, 
but that the color temperature correction stays 
at 1.7,  then it predicts that the intrinsic radius is 
2.89 ($=1.7^2$) times bigger than that given by the {\tt diskbb} normalization.
There can also 
be other corrections to the inner disk radius, depending on how many 
disk photons are intercepted by the Comptonization region and its
geometry with respect to the line of sight 
(Kubota \& Done~2004; Done \& Kubota~2006; Makishima et al.~2008). 
All these corrections act to increase the inner radius, 
so our estimates above using the normalization of {\tt diskbb} are 
a lower limit
to the size of the region contributing the constant component.
For Obs. 1, we can compare the parameters of this constant component
with those derived from spectral fitting of the disk in the time
averaged spectrum in Paper~I. The spectral fitting gives a slightly
higher temperature, hence a lower radius, but just consistent given
the large uncertainties. 
This highlights the difficultly in 
unambiguously separating out the disk component in these spectra. 
The model independent extraction used here is a more robust technique as
it separates out the soft disk component from its lack of variability.
These are consistent with the previous results 
that the disk variability is strongly suppressed above 
$\sim$ 1~Hz (i.e., timescales less than 1~s) while the corona generates variability 
up to $\sim$ 10~Hz (Wilkinson \& Uttley 2009; Uttley et al. 2011). 

The resulting parameters of the constant disk component generally show
a temperature which increases as the source brightens. 
The corresponding uncorrected disk inner radius is 
consistent with radius
decreasing from $100$ to $10R_{\rm{g}}$ with increasing luminosity, but this trend
is not very significant and the data could also be fairly well
described by a constant radius of $\sim 20 R_{\rm{g}}$. 
The former case is clearly
consistent with the truncated disk models, especially as all the
correction factors to the radius act to substantially increase these
estimated values. Even the latter case is significantly larger than
the innermost stable circular orbit of a Schwarzschild black hole, so
is not consistent with an untruncated disk, and especially one which
is strongly illuminated in its central regions, as in the lamppost
reflection models (Fabian et al.~2012). This conflicts with the much
smaller radii derived from the spectral fits of Miller et al.~(2012),
but their adoption of a broken power law continuum gives substantially
more photons in the continuum at the disk energy than a Comptonization
component, suppressing their estimated disk luminosity and hence
leading to a much smaller radius. Our parameters are more robustly
estimated as extracting the constant component from the lightcurves
means that the disk is not dependent on the assumed continuum form. 

\subsection{Comparison of long- and shot-term spectral changes}

We highlight the nature of the spectral changes on long and short
timescales in the hard intermediate state (Obs~14) and relatively dim
low/hard state (Obs~8) in figure~10a and b, respectively. The long
timescale changes (weeks-months: blue points) are shown by the ratio
of each dataset to Obs~3 (taken from figure 6), while the short timescale
(seconds: red points) are the ratio of the high and low spectra from
that observation (taken from figure 8). 

For the hard intermediate state (Obs.~14: figure 10a) the long and short
timescale ratios behave in a drastically different manner below 2~keV.
The short-term variation amplitude steeply {\it
decreases} below 2 keV, down to $\sim 1.2$ at $\sim 0.5$ keV.  As
already argued in section 4.3 and shown in figure 7, this can be
understood as a consequence of the stable disk emission diluting low
energy variability.  
Conversely, the long-term spectral ratio
clearly {\it increases} towards lower energies, reaching an order of
magnitude at 0.5 keV.  We hence conclude that the disk emission, which
lacks fast variations, evolves significantly on long time scales.
Above 2~keV, the two ratios behave in a similar way (except for some
differences in the amplitude).  Both are characterized by 
spectral softening as the source gets brighter. 

The dim low/hard state behavior is rather different (Obs. 8: figure 10b). Both long
and short timescale ratios look very similar, being approximately
constant above $\sim 2$~keV, then rising below this. 
The flatness of the short term ratio above 2~keV shows that
the Comptonization component here varies predominantly in
normalization only, rather than softening as it brightens.

\begin{figure*}[htbp]
    \begin{center}
      \includegraphics[width=0.92\textwidth]{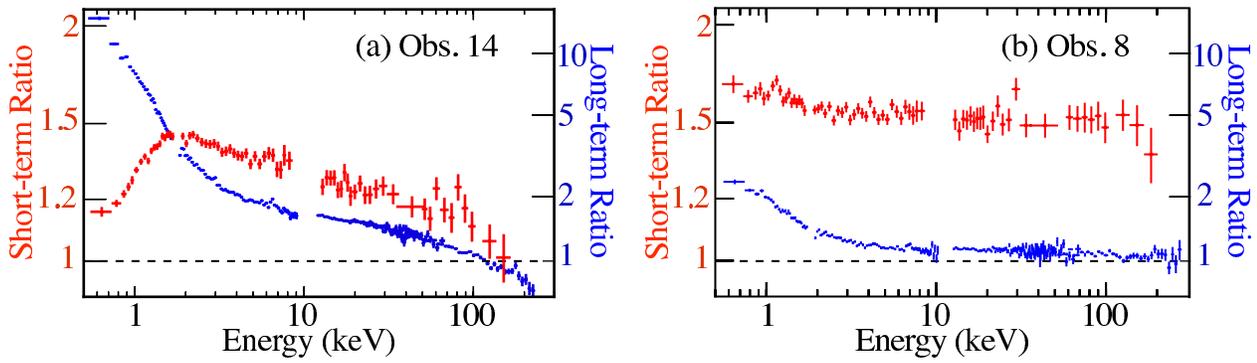}
            \end{center}
    \caption{
    A direct comparison of the long-term spectral ratio (blue) in Obs.~14 as presented in figure \ref{eeufratio},
    and the short-term ratio (red), Obs. 14 vs. Obs. 3, reproduced from figure \ref{hiloratio}. }
    \label{maxcomp}
\end{figure*}

\section{Identifying a Soft Compton Emission} 

The previous sections have clearly identified the main characteristics
of the broad band spectral evolution as an increase in disk, 
correlated with softening of the Compton component, and possibly a decrease in
its electron temperature. Here we explore the more subtle changes 
indicated by the ``bend'' in the spectral ratios in the hard intermediate state spectra 
below $\sim $10~keV. 
This could indicate an additional component beyond
the disk and single temperature Comptonization continuum. 

We constrain the shape of this additional emission in a
model-dependent but rather robust way.  We assume that the hardest
component is a hard thermal Comptonization component ({\tt compps})
and its simple reflection without fluorescence lines which is built into this code, with
ionisation fixed at $0$ since this cannot be constrained from the high
energy data.  We include also the disk component determined from the
{\tt diskbb} fits to the constant component in the previous section.
We use this to fix the seed photon temperature for the Comptonization
if the constant disk component is significantly detected, otherwise we
fix this at 0.1~keV. We fit this model to a high energy bandpass, and
then extrapolate the best-fit spectrum down to lower energies and subtract
this from the data to find the excess emission over and above that
expected from a single thermal Comptonization component with standard
reflection.


\begin{figure*}[htbp]
    \begin{center}
      \includegraphics[width=0.88\textwidth]{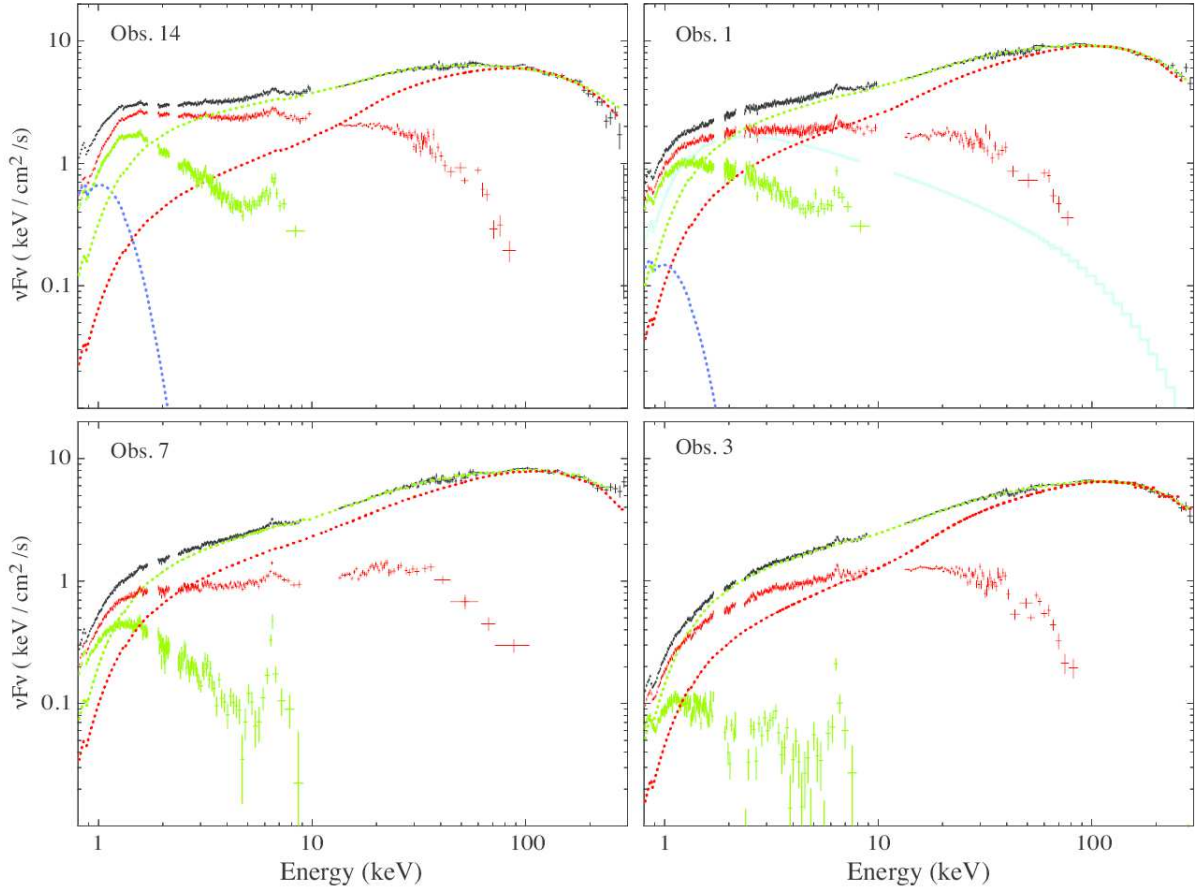}
            \end{center}
    \caption{
   The time-averaged spectra in Obs.~1, 14, 7, and 3 are 
   shown in black (the same as figure 5). 
   Obs. 14, 1, 7, and 3. are the descending of the soft X-ray flux.
   The spectra which are created by 
   subtracting either a hard Compton component including a reflection
   continuum obtained by fitting over 10--300 keV (red dotted curves) or
   over 100--300 keV (green dotted curves), and the raw disk if any (blue dotted curves) 
   from the time-averaged one (black) are shown in the respective color.  
   The soft Compton spectrum used in Paper~I is overlaid in cyan. 
    }
    \label{maxcomp}
\end{figure*}

We first choose the highest energy bandpass; 
i.e., we use only the data 
above 100~keV to determine the properties of the Comptonization
region (red dotted curves in figure~11).  
The red dotted curve shows the model fit with ({\tt compps}) over 100--300 keV, 
and the blue dotted curve is the constant disk if any,   
which are subtracted from the original total spectra shown in black. 
Figure 11 shows the resulting residual emission for Obs. 1, 3, 7
and 14 as the red spectra. 
The solid cyan curve superimposed on Obs.~1 in figure 11 is the
shape of the additional soft Comptonization model derived by full
spectral fitting in Paper~I. Thus this much less model dependent
approach gives results which are consistent with full spectral
modeling. The residual emission (red spectra in figure~11) 
is rather soft with a photon index of $\sim$
2--3, and increases in relative importance with increasing luminosity.
The rollover at high energies is probably not significant since the
residual spectrum must go to zero at 100~keV due to the assumption
that the spectrum above 100~keV contains 
only the hard Comptonization component. 

We contrast this maximal residual emission with the minimum residual
emission, which is that derived from fitting the 10--300 keV PIN+GSO spectra 
with a single Comptonization model and its in-built reflection as in
Paper~II (green dotted curves in figure~11).  
Subtracting this, along with the constant disk component if any (blue dotted curves in figure~11), 
from the total spectrum (black in figure~11) 
results in the green spectrum shown for each observation in figure 11.
This is much softer, as it must go to zero at $\sim 15$~keV, 
but still very significantly present in these spectra.  
Thus, if our assumption is correct,  
the soft Compton emission can be present between 
the red and green spectra in figure~11, 
and it is also likely that 
the soft Compton emission increases as the soft X-ray flux increases. 

We note how the shape and the intensity of the iron line in the residual emission are quite
different depending on the assumed hard component shape. 
Nowak et al.~(2011) quantify this with four low/hard state Suzaku spectra (Obs.~1-4
here) and show that the inner radius derived from the iron line
profile can significantly change with the assumed continuum form.
However, here we are focussing on more robust attributes of the data,
namely the continuum shape.  It is clear from this analysis that there
is more spectral complexity than can be modeled by a disk, single
temperature Comptonization and its reflected emission.  
There is significantly more flux at soft energies
than predicted by these models, and this additional flux increases
with mass accretion rate. 

Miller et al.~(2012) obtained strongly reflection-dominated solutions:
i.e.,  the ratio of reflection to powerlaw flux is larger than 1.0 for 10 out of 20 observations. 
Assuming that the reflection albedo in hard state is $\sim$ 0.2, 
this implies the reflection fraction is larger than $\sim$ 5 (up to $\sim$ 10 in some observations). This is inconsistent with most of previous studies that 
have found the reflection fraction is less than 1.0
 (e.g., Gierli\'{n}ski et al.~1997, Paper~I, and Paper~II).
We additionally note that 
it is important to separate out individual components from the spectrum 
in a model-independent manner, pin down their time evolution, and then reveal their origin,
which may not be a unique way but a reliable method to reach a shred of the truth.

\section{Discussion}

\subsection{Summary of the results}

We analyzed the 25 Suzaku data sets of Cyg X-1 acquired from 2005 to
2009, and successfully obtained high-quality 0.5--300 keV spectra
(figure \ref{alleeuf}).  The source remained in the low/hard and hard
intermediate states throughout, and the unabsorbed 0.1--500 keV luminosity
changed from 0.8--2.8\% of the Eddington limit during the observations.
Dipping periods and piled-up events were properly removed 
(Appendix 1 and 2).

We studied the variations of the 
wide-band spectra on two distinct time scales; 
weeks to months, and 1--2 s. The behavior shows a split between the
brighter low/hard and hard intermediate state spectra, and the dimmer
low/hard state spectra. We list these separately below.

\vspace{0.5cm}
\hspace{-0.3cm}The brighter low/hard and hard intermediate state:

\begin{enumerate} 
\item The fast variability (1--2s) shows a clear separation
between a constant component on a short time scale at low energies and variable tail
extending to high energies. The constant low energy component is
almost certainly the disk and its parameters imply that it is 
almost certainly truncated.

\item The spectra are more complex than
can be represented by a single thermal Compton scattering component
and its simple reflection. Using this model to fit the highest energy
data clearly reveals additional emission with
$\Gamma\sim $2--3 which increases in importance at higher mass
accretion rates.


\vspace{0.5cm}
\hspace{-0.8cm}The dimmer low/hard state:

\item The fast variability (1--2 s) does not show significant drop in $E < 2$ keV, 
rather peaks close to the lower limit of the XIS bandpass. 
The constant low energy component is thought to be rather weaker 
than that in the hard-intermediate state. 
\item The extent of the ``softer when brighter'' on 1--2 s time scale is limited to $E<$3~keV. 
\item In the dimmest low/hard state (Obs.~20 and 4), 
they rather shows the ``harder when brighter'' behavior on weeks or longer time scale. 



\end{enumerate}

While some of these have been suggested by previous instruments, 
they are much more clearly shown by the Suzaku data due to its combination 
of broad bandpass and excellent statistics, 
especially as there are multiple observations 
tracing the behavior as a function of mass accretion rate. 
This is most evident in revealing the soft continuum component, 
which can be seen by eye in the $\nu F_\nu$ form (figure~5) in the hard
intermediate state spectra, but which previously has been
inferred only by detailed (and model dependent) spectral fitting. 
We discuss each of these in more detail below, and then propose a possible geometry 
which can explain the origin of all these components and
their evolution with mass accretion rate.

\subsection{Existence of the constant cool disk}

We detected a sharp decrease of variability in $E \lesssim 2$~keV on
a time scale of $\sim$ 1 s in the brighter low/hard and hard
intermediate states.
We used this to extract the spectrum of the stable component
responsible for the low-energy variability suppression, and reproduced
it with a disk emission model, with a typical temperature of 0.1--0.2~keV
(section \ref{sec:quanhl}).  As listed in table \ref{fitdiskpara}, the
inferred radius for this constant component is generally larger than the
innermost stable circular orbit of a Schwarzchild black hole, 
which is not easily consistent with an untruncated disk.  Conversely, it is
consistent with a truncated disk, and the trend in the data (though
with large uncertainties) is consistent with the truncation radius
decreasing as the mass accretion rate increases as proposed by the
truncated disk models.

These truncated disk models also give a framework in which to
understand the lack of this stable disk component in the lowest
luminosity spectra, as here the disk could be truncated at even larger
radii, so its temperature is too low to significantly contribute to
the spectrum above the lower limit of the XIS bandpass.

We can compare our results with theoretical expectation.  
In the standard disk theory \citep{1973A&A....24..337S}, the viscous time scale, 
$t_{\rm{vis}}$, becomes $\gtrsim$~1~ks at $30$ $R_{\rm g}$, assuming a typical
value of $H/r \sim 0.001$ and a viscous parameter $\alpha \sim 0.1$, 
where $H$ and $r$ are the scale height 
and the radial distance of the disk, respectively. 
On the other hand,
in an optically thin and geometrically thick accretion flow
\citep{1995ApJ...444..231N}, 
we expect $t_{\rm{vis}}$ $\sim$ 70 ms at 
$20$ $R_{\rm g}$, assuming $\alpha \sim 0.1$ and $H/r \sim
1$.  Thus, from a theoretical point of view, 
the disk would vary on $>1$~ks, 
while the corona would change on a time scale of $\lesssim$~1~s.
This agrees with our findings (and also Paper~I) that the disk
emission is more stable than the Comptonization signals.

Considering all these results, we can conclude that a cool and truncated
disk is present in the XIS bandpass in the brighter low/hard state and
hard intermediate state, and is stable on a time scale of $\sim$ 1s. 
Note that this constant disk is seen also in the high/soft state, where
\citet{2001MNRAS.321..759C} showed that it can be distinguished from
the variable powerlaw component on a time scale of $\sim$ 16 s. 

\subsection{Long-term softening: soft Comptonization} 

As presented in Paper~II, the spectra in $ E > 10$ keV in the low/hard
state can be reproduced by a hard Comptonization.  
Furthermore, we have obtained evidence that there is
an additional soft component, and constrained its spectral shape in a
model-dependent but fairly robust way (section 5: figure 11).  

The origin of this component is clearly an important question. 
Paper~I identifies this as an additional soft Comptonization component, 
while Nowak et al.~(2011) suggests that it could be a soft component from a
jet, or that the emission region is homogeneous but with an electron
distribution which is more complex than a single temperature
Maxwellian (see also Ibragimov et al 2005), 
while Fabian et al. (2012) identify this with complex reflection in an extreme 
light-bending scenario. 
However, we can also use the fast variability information
from Paper~II and other studies. 
These clearly show that the soft leads the hard, 
by an amount which depends on frequency of the
variability. For fixed energy bands, slow variability shows a longer
hard lag than fast variability (Nowak et al. 1999; Revnivtsev et al.~1999). This can be interpreted in the context of propagating
fluctuations in an inhomogeneous source, where the spectrum is harder
at smaller radii 
(Kotov et al.~2001; Ar{\'e}valo \& Uttley 2006). 
Independent evidence for an inhomogeneous source is seen by the difference in spectral shape of the fast variability compared to the slower variability, as required
in the models above (Revnivtsev et al.~1999). 
Thus a two Compton component model for the spectrum 
(e.g. Kawabata \& Mineshige 2010) can also produce
the observed complex variability properties. 
A soft jet component clearly has much more difficulty in producing the variability, 
as the fluctuations might be expected to propagate down through the accretion
flow first, then be ejected up the jet, giving a hard lead rather than
the observed lag. 
Similarly, soft emission from reflection should lag 
behind the hard illuminating component variability, 
though this lag time should be very short compared to the measured lags. 
Since a two component Comptonization model can
account for both the spectrum and variability together, we favor this
as the simplest model, and henceforth term the additional soft
component ``the soft Compton component''. 

At low mass accretion rates or in the dim low/hard state, 
the spectral shape above $\sim$ 3 keV remains approximately constant 
on a time scale of 1--2 s.  
The reason is that 
the hard Compton component dominates the spectrum above $\sim 3$~keV, 
varying mainly with its normalization; 
in other words, 
the soft Compton component in the dim low/hard state 
is much weaker relative to the hard Compton component. 
Conversely, it is the
increasing dominance of the soft Compton component in the hard
intermediate state which leads to the abrupt softening on the
color-intensity diagram which is the defining characteristic of this
state (figure 2b). 
These features suggest that 
the soft Comptonization component probably has different time variability,
characterized by ``softer when brighter''.

Therefore, the present results significantly strengthen the idea of multi-zone
Comptonization presented in Paper~I.  
A new implication obtained here is 
that the soft Comptonization changes more than the hard
Comptonization on the long time scale, 
possibly following more faithfully the accretion-rate changes.

\subsection{Where do soft seed photons come from?} 

\begin{figure}[htbp] 
    \begin{center} 
      \includegraphics[width=0.48\textwidth]{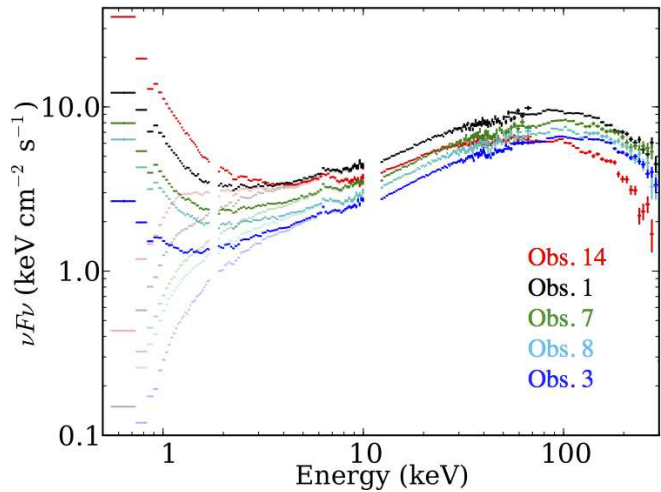} 
            \end{center} 
    \caption{ 
   Several representative $\nu$$F$$_\nu$  spectra after removing interstellar absorption. In comparison, 
   those before removing  interstellar absorption are shown in faint colors. 
       } 
    \label{noabs_spec} 
\end{figure}

In the hard intermediate state, 
the spectra become softer and show decrease of fast variability in $E<2$ keV 
as shown in figure~8 when it gets brighter. 
This trend is the same as found in Paper~I, 
and thus the interpretation proposed in Paper~I  
that the seed photons are put into the inhomogeneous corona above the disk can be valid. 
On the other hand, the break energy in the spectral ratios on a time scale of $\sim$ 1 s in figure~8, 
decreases from $\sim$ 2 keV in the hard intermediate state 
to $<$ 0.5 keV in the dim low/hard state, 
as the soft X-ray flux decreases. 
The new observational fact requires supplemental explanation 
to the physical interpretation proposed in Paper~I.  
This might appear to be simply explained by 
the increase of the truncation radius and the decrease of the disk temperature. 
However, 
the soft excess in $E < 2$~keV can be clearly recognized in figure \ref{noabs_spec} 
by looking into the time-averaged $\nu$$F$$_\nu$ spectra 
after removing the typical interstellar absorption of $N_{H} = 7 \times 10^{21}$ cm$^2$. 
Furthermore, 
the variability in $E < 2 $ keV does not decrease, 
but rather increases (Obs.~3, 20, and 4 in figure~8). 
Therefore, the spectrum should include some extent of variable soft emission, 
which can not be separated out on 1--2 s time scale as a constant disk component. 
This is consistent with a soft component inferred from 
detailed spectral fits to the low/hard state data; 
e.g., according to Nowak et. al (2011), 
the spectra in Obs. 2 and 3 show a clear need for a thermal component at low energies, 
yet this is not seen as a constant component in our analysis.

The variable soft component, if its spectral shape is similar to disk emission, 
would probably have a slightly higher temperature than 
the (unseen) constant disk component, 
because the disk temperature should be higher than $\sim$ 0.1 keV 
to create soft excess up to $\sim 1 $ keV. 
The simplest interpretation within a framework of the ``disk-corona'' configuration would be 
that the disk interacts more strongly with the corona 
in the dim low/hard state than in the hard-intermediate state, 
leaving less stable disk-like component in the dim low/hard state, 
but providing some amount of seed photons for the corona in a time-varying way, 
presumably at an overlapping region between the disk and the corona. 
The slight ``softer when brighter'' behavior 
around 0.5--2 keV in the dim low/hard state 
might be some hint of variable soft emission 
if it is indeed produced by Comptonization (Gierli\'{n}ski \& Zdziarski 2004). 

Alternatively,  
the source of seed photons might be different when accretion rate is low, 
e.g., synchrotron emission (Wardzi{\'n}ski \& Zdziarski 2000; Veledina et al. 2011a,b). 
In addition to the different short-term behavior in $E < 2$ keV, 
the faintest Obs. 3, 20, and 4 with $L \sim$ 1\% $L_{\rm{Edd}}$
exhibit somewhat anomalous long-term
spectral changes over the entire 0.5-200 keV range; the spectrum {\it
softens} as the source gets dimmer (figure 6).  This behavior is
commonly seen in other  black hole binaries 
at $L \lesssim$ 1\% $ L_{\rm{Edd}}$ (Yamaoka et al. 2005; Yuan et al. 2007; Wu
\& Gu 2008; Russell et al. 2010; Sobolewska et al. 2011; Armas Padilla et al. 2013). 
The ratio fluxes during our observations taken in 2009 are 
monitored with the Arcminute MicroKelvin Array radio telescope (Zwart et al.~2008). 
The 12--70 keV X-ray and ratio fluxes are positively correlated, 
as shown in figure 4 in Miller~et al. (2012). 
Thus, it does not indicate the increase of synchrotron radiation, 
but imply the possibility 
that seed photons start to be more provided by cyclo-synchrotron radiation 
than the truncated stable disk 
when $L \lesssim$ 1\% $L_{\rm{Edd}}$ (Sobolewska et al. 2011; Gardner \& Done 2012). 




\subsection{Disk and Inhomogeneous Coronae configuration} 

We can put all these together if the spectrum contains a constant disk, 
a variable soft Compton component which softens as it brightens
on short timescales, 
and a variable hard Compton component which
changes mostly in normalization, 
and possibly the additional variable disk-like component. 
As found in the dim low/hard state, 
the disk is truncated far from the black hole, 
so its temperature is below the lower end of the XIS bandpass. 
Increasing the mass accretion rate gives an increasing
temperature and luminosity from the constant disk, increasing its
impact on the XIS spectrum. This produces the characteristic drop in
fast variability at low energies seen increasingly at higher
luminosities. 

The fraction of the flow in the soft Comptonization regime, 
which is presumably generated from overlapping region 
between the disk and corona, 
increases with mass accretion rate. 
The soft Compton comes to dominate 
the spectrum below 10~keV in the hard intermediate state, 
but only makes a much smaller contribution, 
extending only to $\sim$ 2~keV in the dim low/hard state. 
Fast variability supports this, 
because the spectrum is ``softer when brighter'' in the hard intermediate state 
while almost no change in shape in the dimmer low/hard state. 

\begin{figure*}[htbp]
    \begin{center}
      \includegraphics[width=0.95\textwidth]{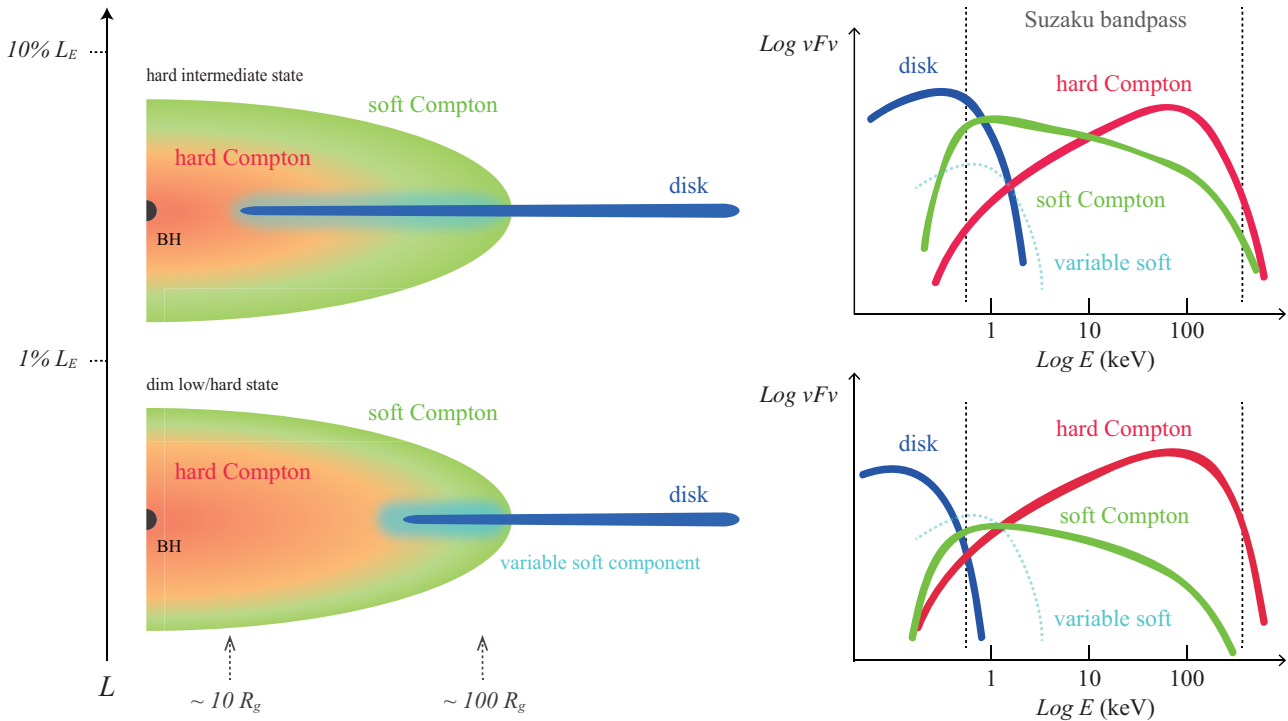}
            \end{center}
    \caption{
    Conceivable accretion flows (left) and spectral composition in $\nu F_{\nu}$ form (right)
    in the hard intermediate state (top) and the dim low/hard state (bottom). 
    Some gross representative values on luminosities and radii are overlaid on the flow pictures.
    The disk, hard Compton, soft Compton, and variable soft component are indicated by blue, red, green, and cyan, respectively. Fe-K lines and reflection components, 
    which presumably come from the disk, are not depicted just for simplicity.  
    }
    \label{disc_pic}
\end{figure*}

We sketch this changing geometry in figure \ref{disc_pic}.  We
propose that this is a universal geometry for the low/hard and hard
intermediate states as wide-band Suzaku spectra of GRO~J1655-40
\citep{2008PASJ...60S..69T} and GX~339-4 \citep{Shidatsu2011} were 
also well reproduced with the double-Compton modeling, which includes a 
disk with $kT_{\rm{in}} \sim 0.2$ keV. 


As mentioned in section 6.4, 
the constant truncated disk is likely to be accompanied by a variable soft component, 
which might be originated from highly unstable disk-corona interaction or   
a variable cyclo-synchrotron radiation or else. 
It would be possible to speculate perhaps 
clumps torn off the disk edge (Chiang et al. 2010). 
These clumps would be variable, 
as they are continually removed 
by being accreted and/or dissipated/evaporated,
but are also continually replenished by new clumps forming. 
They are likely to have a higher temperature, and smaller emitting area 
than the constant disk component. 
Thus they are still visible in the spectrum when the constant disk has too low a temperature to
contribute to the observed bandpass, 
as in the dim low/hard state. 
These would also be closer (embedded into?) the Comptonising region, 
so that they can efficiently put seed photons for the Comptonization. 


\subsection{Analogies to massive black holes}

The primary emission from active galactic nuclei, AGNs, 
which is also regarded as Comptonization based on the analogy of the black hole binaries, 
should be understood before decoding reprocessed or secondary emission in detail. 
Recently, Suzaku spectra of the typical Seyfart galaxy Mrk 509 have
been reproduced by two-component Comptonization (Noda et al. 2011b;
Mehdipour et al. 2011).  Noda et al. (2011a) analyzed Suzaku data of
MCG--6-30-15, and found a secondary component that is different in
variation properties from the dominant powerlaw. 
Similar components are reported in different types of AGNs (Noda et al. 2013). 
A more likely contender for similar behavior is the 
typical low-luminosity AGN; e.g., 
``softer when brighter'' behavior is seen in NGC~4258 (Yamada et al 2009), 
while ``harder when brighter'' in NGC~7213 (Emmanoulopoulos et al. 2012). 
Although these samples are not enough, future missions, which have
wider spectral coverage with higher sensitivity, such as ASTRO-H
(Takahashi et al. 2010), enable us to reveal the nature of the
disk-coronae picture around a massive black hole.

\section{Conclusion}

We investigate the spectral evolution via model-independent analysis
of the long and short timescale variability. 
Our results are summarized as follows: 

\begin{itemize}
\item A cool disk component exists almost certainly in the low/hard state, 
which increases in luminosity and temperature as the luminosity increases.
The disk parameters imply that it is truncated, and are consistent
with radius decreasing as the luminosity increases. 
\item The bright low/hard state spectra show a clear break at $\sim 10$~keV 
on long time scales, as opposed to the lower luminosity spectra, where the
3-300~keV tail is consistent with a single Compton component (and its
simple reflection). This can be interpreted as two separate Compton components, 
one hard and the other soft Compton component as obtained in Paper~I. 
\item On long timescales, the soft Compton component increases along with the cool disk component, 
while the hard Compton does not change by much at all. 
Thus the hard intermediate spectra are dominated by the soft
Compton component at least up to $\sim 3$--$30$~keV, 
with the hard Compton component 
dominating at the highest energies, while the dim low/hard state
is dominated by the hard Compton component from 3--300~keV. 
\item An additional, variable, source of seed photons is implied 
in the dim low/hard state based on the lack of decrease in variability as shown in figure 8 
and the variable soft excess as shown in figure~12. 
Conceivable accretion flows are presented in figure~13. 
\end{itemize}

%


Thus there can be potentially four components in the Suzaku bandpass: a
cool truncated disk which is constant, a variable soft blackbody-like component, 
a soft Compton component from the outer flow and a hard
Compton component from the inner flow. Spectral fitting alone is
unlikely to uniquely separate out all these component and we urge a
combined spectral-timing approach in order to robustly interpret such
complex data.

\section*{Acknowledgement}
The authors would like to express their thanks to Suzaku team members. 
The research presented in this paper has been financed by
Grant-in-Aid for JSPS Fellows. 
S.Y. is supported by the Special Postdoctoral
Researchers Program in RIKEN. This work was supported by JSPS KAKENHI Grant Number 24740129.

\appendix

\section{Orbital phase and exclusion of dipping periods}
\label{sec:dip}

\begin{figure}[htbp]
    \begin{center}
      \includegraphics[width=0.48\textwidth]{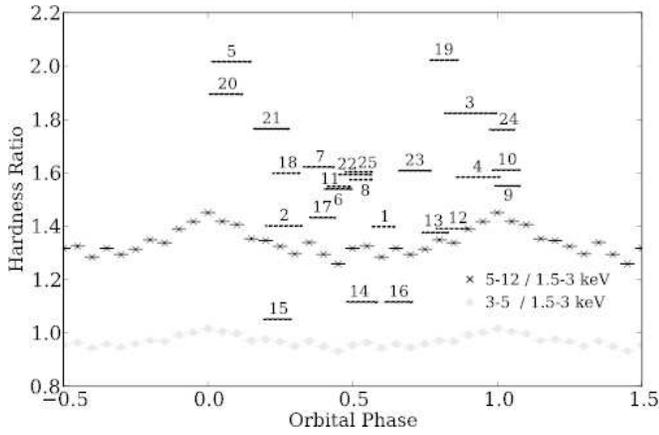}
            \end{center}
    \caption{The hardness ratio of ASM of Cyg X-1 folded by the orbital period of 5.6 days. Two phases are shown for clarity.  Phase 0 is defined as the superior conjunction, when the black hole is the farthest from us. The corresponding Suzaku observations are indicated, with their vertical positions representing the 5--12/1.5--3 keV hardness ratio. The phase coverage by each Suzaku observation is indicated by the length of each segment.} 
    \label{rxtephase}
\end{figure}

In table \ref{obslog}, 
we summarize the orbital phases of all Suzaku observations.
Since photo absorption caused by dips significantly affects soft X-rays, 
we need to property exclude the dip phases. 
We have folded the hardness ratios (table~1) 
of ASM according to the orbital period,  
where data points were discarded while the object is in the high/soft state. 
The orbital period $P = 5.599829$ days and an epoch of the superior conjunction 
of the black hole at MJD$41874.207$ are used, based on Brocksopp et al. (1999). 
Figure \ref{rxtephase} shows the results of this analysis. 
At the near phase 0 (i.e., when the observer, the companion star, and Cyg X-1 are in line with this order), 
the hardness ratio increases by $\sim$ 20\% due to increased absorption.  
This modulation has been studied by many authors (e.g., \cite{Poutanen2008}).

We overlaid the hardness ratios at the 25 observations in figure \ref{rxtephase}. 
Our observations are thus evenly distributed over the orbital phase. 
The scatter of the hardness ratio 
is much larger than that caused by the orbital modulation, 
so that the effect seen among the Suzaku observations 
can be considered as a result of intrinsic spectral changes of Cyg X-1.  

\begin{figure*}[htbp]
    \begin{center}
      \includegraphics[width=0.7\textwidth]{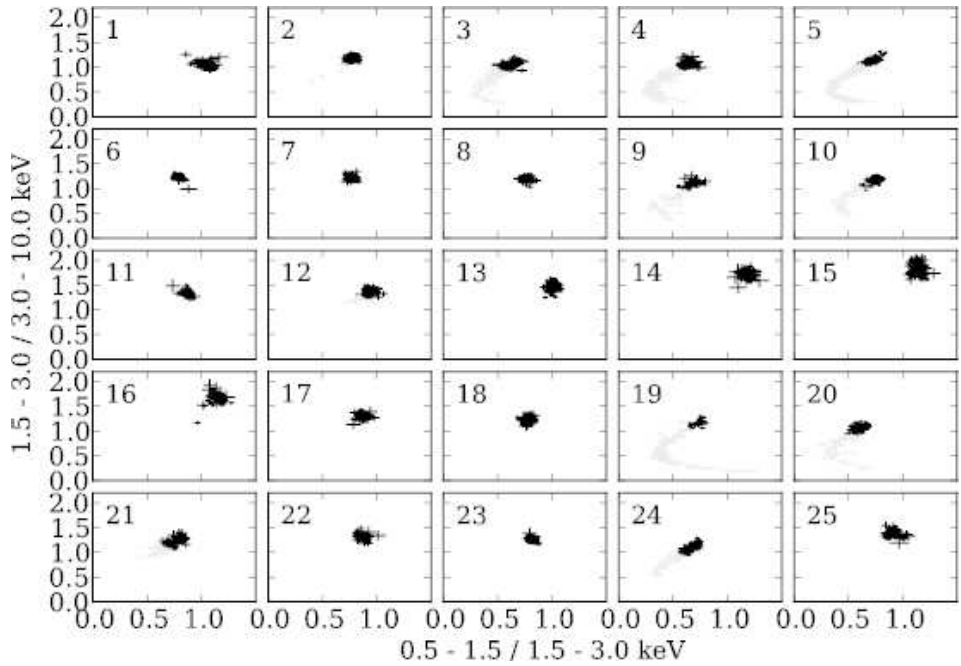}
            \end{center}
    \caption{The 0.5--1.5 to 1.5--3.0 keV softness vs.  the 1.5--3.0 to 3.0-10.0 keV softness plot, for 128 s binning. The grey data points define the dipping period, while the black ones represent normal periods. } 
    \label{sscolor}
\end{figure*}

As seen in figure \ref{rxtephase}, 
some of the observations are observed around phase 0., 
which are likely suffered from significant dips. 
Thus, we need to exclude the dipping period during observation.
We constructed three-band (0.5--1.5, 1.5--3.0, and 3.0--10.0 keV) light curves 
for the same XIS used in figure~3.
These light curves were utilized for studying the source motion on a ``softness-softness plot''  in figure \ref{sscolor},
in which the ratio of 1.5--3.0 keV to 3.0--10.0 keV count rate plotted against that of 0.5--1.5 keV to 1.5--3.0 keV count rate. The dips start with an increase of the neutral absorption, 
followed by an enhanced contribution by an ionized or partial-covering absorber. 
The former effect brings the data points toward lower left, 
while the latter toward lower right due to recovery of the softest counts. 
These dips last for several hours, causing significant drops in the soft X-ray flux. 
It has already been reported 
that neutral column density changed from $\sim 5 \times 10^{21}$ cm$^{-2}$ to $\sim 5 \times 10^{22}$ cm$^{-2}$ in Obs. 3, 4, and 5 (Nowak et al. 2011).

We define ``dipping periods'', as those time bins 
wherein the data on the softness plot is outside a circle 
with a radius of 0.2, centered on the distribution centroid. 
These periods are indicated in light gray in figure \ref{sscolor}. 
Throughout the paper, we used the period excluding the dipping phase. 

\section{Attitude correction and pileup estimation of the XIS data}
\label{sec:att-pl}

\begin{table*}[htbp]
\caption{The exposures, estimated $R_{\rm3\%}$ and $R_{\rm1\%}$,  and corresponding count rates for each XIS.}
\label{xisinfotbl}
\begin{center}
\begin{tabular}{crrrrrrrrr}
\hline\hline
N & \multicolumn{3}{c}{XIS0} & \multicolumn{3}{c}{XIS1}  & \multicolumn{3}{c}{XIS3}  \\[2mm] 
            & $E$$^{*}$(ks) & $R$$^\dagger$$_{\rm 3,  1\%}$ & $C$$^\ddagger$$_{\rm a,  3,  1\%}$  & $E$(ks) & $R$$_{\rm 3,  1\%}$ & $C$$_{\rm a,  3,  1\%}$  &  $E$(ks) & $R$$_{\rm 3, 1\%}$ & $C$$_{\rm a,  3, 1\%}$  \\[2mm]                      
\hline
1$^\S$ & 10.5 & 74, 109 & 280, 67, 18 & 3.5 & 56, 127 & 241, 124, 20 & 6.9 & 70, 135 & 298, 101, 23 \\
2$^\|$ & 8.3 & 21, 50 & 216, 163, 77 & 8.2 & 17, 50 & 233, 206, 116 & 8.3 & 23, 51 & 246, 182, 93 \\
3 & 11.3 & 18, 49 & 145, 122, 74 & 11.3 & 11, 48 & 156, 149, 102 & 11.3 & 21, 53 & 168, 139, 86 \\
4 & 8.3 &  8, 35 & 95, 89, 61 & 8.3 &  6, 32 & 102, 100, 78 & 8.3 & 10, 38 & 109, 102, 70 \\
5 & - & - & - & 8.5 & 19, 56 & 197, 175, 112 & 8.5 & 25, 61 & 205, 160, 91 \\
6 & 3.8 & 31, 71 & 264, 186, 88 & 3.8 & 30, 79 & 299, 240, 116 & 3.8 & 36, 75 & 303, 203, 99 \\
7 & 5.3 & 30, 70 & 250, 180, 86 & 5.3 & 27, 78 & 273, 227, 108 & 5.3 & 34, 74 & 285, 201, 98 \\
8 & 4.3 & 27, 62 & 212, 160, 87 & 4.3 & 24, 72 & 240, 205, 107 & 4.3 & 30, 68 & 242, 178, 93 \\
9 & - & - & - & 4.5 & 18, 57 & 203, 185, 119 & - & - & - \\
10 & 5.1 & 27, 61 & 214, 160, 87 & 5.1 & 22, 69 & 233, 205, 112 & 5.1 & 31, 68 & 252, 184, 97 \\
11 & 4.9 & 35, 74 & 300, 203, 95 & 4.8 & 36, 89 & 352, 270, 119 & 4.9 & 39, 82 & 353, 235, 109 \\
12 & 5.3 & 34, 72 & 282, 192, 94 & 5.3 & 36, 87 & 336, 256, 114 & 5.3 & 38, 78 & 333, 222, 109 \\
13 & 5.2 & 44, 84 & 385, 230, 99 & 5.1 & 46, 101 & 448, 311, 109 & 5.0 & 50, 90 & 447, 254, 115 \\
14 & 6.9 & 50, 90 & 429, 235, 99 & 6.1 & 51, 105 & 496, 331, 117 & 5.5 & 56, 94 & 479, 251, 116 \\
15 & 3.6 & 61, 99 & 522, 242, 100 & 2.8 & 62, 111 & 597, 342, 127 & 2.7 & 62, 103 & 565, 269, 115 \\
16 & 4.2 & 46, 88 & 418, 242, 99 & 4.1 & 49, 101 & 484, 327, 128 & 4.0 & 55, 92 & 479, 251, 118 \\
17 & 5.1 & 29, 65 & 226, 167, 89 & 5.1 & 27, 75 & 261, 215, 108 & 5.1 & 32, 71 & 258, 183, 93 \\
18 & 5.8 & 29, 64 & 221, 163, 88 & 5.8 & 27, 74 & 256, 211, 108 & 5.8 & 31, 69 & 248, 179, 92 \\
19 & 5.3 & 10, 40 & 112, 105, 68 & 5.6 &  7, 40 & 129, 125, 90 & 5.7 & 14, 44 & 131, 117, 76 \\
20 & 6.5 & 16, 47 & 135, 119, 75 & 6.5 & 10, 44 & 143, 137, 95 & 6.5 & 18, 50 & 150, 128, 78 \\
21 & 5.8 & 34, 72 & 280, 193, 95 & 5.9 & 34, 85 & 317, 243, 109 & 6.0 & 37, 76 & 324, 214, 103 \\
22 & 5.4 & 33, 71 & 272, 189, 95 & 5.6 & 34, 84 & 315, 240, 109 & 5.6 & 37, 74 & 309, 205, 101 \\
23 & 5.5 & 33, 67 & 260, 182, 97 & 5.6 & 32, 84 & 297, 232, 104 & 5.6 & 36, 72 & 294, 198, 102 \\
24 & 4.7 & 26, 62 & 192, 153, 83 & 5.4 & 24, 65 & 225, 194, 112 & - & - & - \\
25 & 2.7 & 48, 90 & 407, 232, 95 & 5.4 & 51, 96 & 457, 281, 114 & - & - & - \\
\hline\hline
\end{tabular}
\end{center}
\begin{itemize}
\setlength{\parskip}{0cm} %
\setlength{\itemsep}{0cm} %
\item[$*$] The exposure after removing the periods of telemetry saturation. 
\item[$\dagger$] In units of pixel, where 1 pixel is 1.042$''$.
\item[$\ddagger$] The count rates integrated over a circle with a radius of 4$'$, and outside $R_{\rm3\%}$ and $R_{\rm1\%}$. 
\item[$\S$]  $R_{\rm3\%}$ and $R_{\rm1\%}$ of XIS2 are 87, 125 pixel, 11.2 ks,  292.0, 56.6, 18.2 counts s$^{-1}$. 
\item[$\|$]  $R_{\rm3\%}$ and $R_{\rm1\%}$ of XIS2 are 23, 53 pixel, 8.3 ks, 257.1, 196.0, and 97.4 counts s$^{-1}$. 
\end{itemize}
\end{table*}

Due to thermal wobbling of the XRT, 
the center of the images fluctuates by $\sim1'$,
depending on day or night for the satellite, 
on a time scale of $\sim$ 45 minutes (half the orbital period of the satellite).   
Since this affects significantly the fraction of the photons \citep{Uchiyama2008}, 
the 1/4 window option has been used since Obs.~3. 
A correction software, {\tt aeattcor} \citep{Uchiyama2008}, has
been developed to correct the attitude solution for the thermal wobbling.  
However, even after processed with this software, 
there still remains attitude fluctuation by $\sim$ 30$''$. 
We further corrected the image for the residual fluctuation, 
and reduced them to be $\sim$ 5 pixel ($\sim5''$).  
More details on the attitude correction are summarized in Yamada et al. (2012). 
%

In general, the extent of the pileup effect is proportional to the incoming flux squared, 
because the phenomenon is a kind of process of two-particle collision. 
The process is irreversible, and thus there is no straightforward ways to estimate the pileup effects. 
We have utilized several phenomenological approaches for bright point sources to estimate the pileup effects; 
e.g., surface brightness, hardness, and grade branching ratio. 

Based on the analyses, 
we compiled a pileup estimation software, {\tt aepileupcheckup.py} 
which returns the radii at which pileup fraction (cf. eq (1) in \cite{2012PL}) becomes
3\% and 1\%.
Table~\ref{xisinfotbl} summarizes the exposure excluding the telemetry-saturated period,   
the two radii corresponding to $R_{\rm3\%}$ and $R_{\rm1\%}$,  
the 0.5--10.0 keV count rates for a circle within a radius of  4$'$, 
outside $R_{\rm3\%}$ and $R_{\rm1\%}$. 
The differences in exposure among the XIS cameras are 
due to differences of their allocated telemetry quota and the employed editing modes.  


\section{Count rates of the HXD data}
\label{sec:hxdcr}

Table \ref{hxdinfotbl} gives the exposures of the HXD data 
corrected for dead time, and NXB-subtracted count rates 
in four energy bands. 
The dead time fraction was estimated by using ``pseudo'' events (Kokubun et al.~2007). 
The observed count rates in Obs.~1 are apparently higher by $\sim$ 10 \% than the others, 
because the effective area is higher by $\sim$ 10 \% in the HXD nominal position employed in that observation than in the XIS nominal position. 

The lower threshold of PIN was gradually raised to avoid enhanced noise since Obs.~6, 
which reduces the effective area of the PIN over 10--15 keV. 
Thus, the count rates in $E > 15$ keV are shown in table \ref{hxdinfotbl}.

\begin{table*}[htbp]
\caption{The exposures and count rates of Cyg X-1 recorded with the HXD.}
\label{hxdinfotbl}
\begin{center}
\begin{tabular}{crrrrrc}
\hline\hline
N & \multicolumn{1}{c}{Exposure} & \multicolumn{2}{c}{PIN}  & \multicolumn{2}{c}{GSO}  \\[2mm] 
                       & Time(ks)$^*$ & $C$(15--20 keV)$^\dagger$  & $C$(20--60 keV)$^\dagger$ & $C$(50--100 keV)$^\dagger$ & $C$(100--200 keV)$^\dagger$  & $F_{10-200}$$^\ddagger$ \\[2mm]                      
\hline
1 & 16.8 (92.9) & 12.61$\pm$0.03 (1.0) & 15.06$\pm$0.03 (1.2) & 17.30$\pm$0.04 (19.0) & 8.94$\pm$0.03 (36.2) & 4.14\\
2 & 25.8 (93.1) & 11.91$\pm$0.02 (1.1) & 14.24$\pm$0.02 (1.4) & 17.67$\pm$0.03 (26.4) & 9.63$\pm$0.03 (46.5) & 3.88\\
3 & 37.4 (93.0) & 8.39$\pm$0.02 (1.4) & 10.07$\pm$0.02 (1.9) & 12.61$\pm$0.03 (34.4) & 7.17$\pm$0.03 (56.2)  & 2.80\\
4 & 30.4 (93.4) & 5.33$\pm$0.01 (2.2) & 6.27$\pm$0.01 (3.0) & 7.75$\pm$0.03 (45.2) & 4.39$\pm$0.03 (67.0) & 1.73\\
5 & 27.0 (93.3) & 10.06$\pm$0.02 (1.1) & 12.05$\pm$0.02 (1.6) & 14.49$\pm$0.03 (32.2) & 7.88$\pm$0.03 (56.0) & 3.24\\
6 & 14.4 (92.7) & 11.32$\pm$0.03 (0.9) & 13.58$\pm$0.03 (1.2) & 16.62$\pm$0.05 (30.5) & 9.06$\pm$0.05 (54.1) & 3.70\\
7 & 12.2 (92.9) & 10.82$\pm$0.03 (1.1) & 13.13$\pm$0.03 (1.4) & 15.89$\pm$0.05 (31.9) & 8.81$\pm$0.05 (55.1) & 3.56\\
8 & 12.2 (93.6) & 9.60$\pm$0.03 (1.1) & 11.60$\pm$0.03 (1.4) & 14.10$\pm$0.05 (33.3) & 7.80$\pm$0.05 (57.3)  & 3.15\\
9 & 15.2 (93.3) & 10.16$\pm$0.03 (1.1) & 12.25$\pm$0.03 (1.4) & 14.70$\pm$0.04 (33.1) & 8.14$\pm$0.04 (57.4) & 3.30\\
10 & 11.4 (91.8) & 9.72$\pm$0.03 (1.2) & 11.58$\pm$0.03 (1.6) & 13.97$\pm$0.05 (36.0) & 7.62$\pm$0.05 (59.2) & 3.14\\
11 & 14.0 (91.5) & 11.24$\pm$0.03 (1.0) & 13.48$\pm$0.03 (1.3) & 15.74$\pm$0.05 (33.3) & 8.34$\pm$0.05 (57.3) & 3.54\\
12 & 16.2 (92.6) & 10.04$\pm$0.03 (1.1) & 11.75$\pm$0.03 (1.5) & 13.26$\pm$0.04 (35.5) & 6.92$\pm$0.04 (60.6) & 3.02\\
13 & 15.0 (92.7) & 12.42$\pm$0.03 (0.9) & 14.44$\pm$0.03 (1.2) & 15.93$\pm$0.05 (31.8) & 7.72$\pm$0.04 (58.7) & 3.58\\
14 & 24.2 (94.1) & 11.30$\pm$0.02 (0.9) & 12.89$\pm$0.02 (1.3) & 13.12$\pm$0.03 (35.0) & 6.04$\pm$0.03 (64.2)  & 3.03\\
15 & 14.1 (93.9) & 14.22$\pm$0.03 (0.8) & 15.84$\pm$0.03 (1.1) & 15.40$\pm$0.05 (31.7) & 6.88$\pm$0.04 (61.0)  & 3.63\\
16 & 6.2 (92.2) & 11.44$\pm$0.04 (1.0) & 13.02$\pm$0.05 (1.3) & 13.56$\pm$0.07 (35.3) & 6.51$\pm$0.07 (62.5)    & 3.14\\
17 & 14.2 (93.6) & 8.81$\pm$0.03 (1.2) & 10.41$\pm$0.03 (1.7) & 12.08$\pm$0.04 (38.2) & 6.45$\pm$0.04 (63.3)     & 2.73\\
18 & 18.6 (93.0) & 9.17$\pm$0.02 (1.2) & 10.98$\pm$0.02 (1.6) & 12.93$\pm$0.04 (37.1) & 7.16$\pm$0.04 (60.9)    & 2.93\\
19 & 12.9 (91.8) & 7.20$\pm$0.02 (1.7) & 8.62$\pm$0.03 (2.2) & 10.29$\pm$0.05 (43.6) & 5.72$\pm$0.05 (66.8)      & 2.33\\
20 & 16.8 (92.3) & 7.08$\pm$0.02 (1.5) & 8.49$\pm$0.02 (2.0) & 10.08$\pm$0.04 (43.5) & 5.64$\pm$0.04 (67.2)      & 2.28\\
21 & 17.4 (93.2) & 12.26$\pm$0.03 (0.8) & 14.91$\pm$0.03 (1.1) & 17.56$\pm$0.04 (29.9) & 9.42$\pm$0.04 (54.1) & 3.93\\
22 & 10.5 (92.1) & 10.36$\pm$0.03 (1.1) & 12.25$\pm$0.03 (1.4) & 14.25$\pm$0.05 (35.4) & 7.77$\pm$0.05 (59.3) & 3.24\\
23 & 12.1 (92.2) & 10.29$\pm$0.03 (1.1) & 12.34$\pm$0.03 (1.4) & 14.71$\pm$0.05 (34.7) & 8.00$\pm$0.05 (58.6) & 3.30\\
24 & 16.4 (94.1) & 10.08$\pm$0.02 (0.9) & 12.07$\pm$0.03 (1.2) & 14.48$\pm$0.04 (33.1) & 8.05$\pm$0.04 (57.8) & 3.26\\
25 & 3.2 (92.4) & 15.57$\pm$0.07 (0.7) & 18.56$\pm$0.08 (1.0) & 21.20$\pm$0.11 (26.7) & 10.95$\pm$0.10 (50.5) & 4.76\\
\hline\hline
\end{tabular}
\end{center}
\begin{itemize}
\setlength{\parskip}{0cm} %
\setlength{\itemsep}{0cm} %
\item[$^*$]  The deadtime-corrected exposures and the live-time fractions in percent in the parentheses. 
\item[$^\dagger$] The NXB-subtracted count rates and the NXB fractions in percent in the parentheses. The errors are statistical errors only.  
\item[$^\ddagger$] The energy flux of $10$--$200$ keV in an unit of 10$^{-8}$ erg s$^{-1}$ cm$^{-2}$.
\end{itemize}
\end{table*}


{}


\begin{thebibliography}{}


\bibitem[Ar{\'e}valo \& Uttley(2006)]{2006MNRAS.367..801A} Ar{\'e}valo, P., \& Uttley, P.\ 2006, \mnras, 367, 801 

\bibitem[Armas Padilla et al.(2013)]{2013MNRAS.428.3083A} Armas Padilla, M., Degenaar, N., Russell, D.~M., \& Wijnands, R.\ 2013, \mnras, 428, 3083 


\bibitem[Beloborodov et al.(1999)]{Beloborodov1999} Beloborodov, A. M., 1999, \apj, 510L, 123


\bibitem[Brocksopp et al.(1999)]{Brocksopp1999} Brocksopp, C., Fender, R.~P., Larionov, V., et al.\ 1999, \mnras, 309, 1063 

\bibitem[Caballero-Nieves et al.(2009)]{Caballero-Nieves2009} Caballero-Nieves, S. M., 2009, \apj, 701, 1895


\bibitem[Chiang et al.(2010)]{Chiang2010} Chiang, C.~Y., Done, C., Still, M., \& Godet, O.\ 2010, \mnras, 403, 1102 

\bibitem[Churazov et al.(2001)]{2001MNRAS.321..759C}Churazov, E., Gilfanov, M., \& Revnivtsev, M.\ 2001, \mnras, 321, 759

\bibitem[Dotani et al.(1997)]{Dotani1997}Dotani, T., et al. 1997, ApJ, 485, L87 

\bibitem[Di Salvo et al.(2001)]{2001ApJ...547.1024D}Di Salvo, T., Done, C., {\.Z}ycki, P.~T., Burderi, L., \& Robba, N.~R.\ 2001, \apj, 547, 1024

\bibitem[Done \& Kubota(2006)]{2006MNRAS.371.1216D} Done, C., \& Kubota, A.\ 2006, \mnras, 371, 1216 
\bibitem[Done et al.(2007)]{2007A&ARv..15....1D} Done, C., Gierli\'{n}ski, M., \& Kubota, A.\ 2007, \aapr, 15, 1 

\bibitem[Emmanoulopoulos et al.(2012)]{2012MNRAS.424.1327E} Emmanoulopoulos, D., Papadakis, I.~E., McHardy, I.~M., et al.\ 2012, \mnras, 424, 1327 


\bibitem[Fabian et al.(2012)]{2012Fabian}Fabian, A. C. et al., 2012, MNRAS, 424, 217

\bibitem[Fender et al.(2004)]{2004Fender} Fender, R.~P., Belloni, T.~M., \& Gallo, E.\ 2004, \mnras, 355, 1105 

\bibitem[Fender et al.(2006)]{2006Fender} Fender, R.~P., Stirling, A.~M., Spencer, R.~E., et al.\ 2006, \mnras, 369, 603 

\bibitem[Frontera et al.(2001)]{2001ApJ...546.1027F} Frontera, F., et al.\ 2001, \apj, 546, 1027

\bibitem[Fukazawa et al.(2009)]{2009PASJ...61S..17F} Fukazawa, Y., et al.\ 2009, \pasj, 61, 17 

\bibitem[Gardner \& Done(2012)]{2012Gardner} Gardner, E., \& Done, C.\ 2012, MNRAS, $in press$ 

\bibitem[Gierli\'{n}ski et al.(1997)]{1997MNRAS.288..958G} Gierli\'{n}ski, M., Zdziarski, A.~A., Done, C., et al.\ 1997, \mnras, 288, 958 


\bibitem[Gierli{\'n}ski et al.(1999)]{Gie1999} Gierli{\'n}ski, M., Zdziarski, A.~A., Poutanen, J., et al.\ 1999, \mnras, 309, 496 

\bibitem[Gierli\'{n}ski \& Zdziarski(2005)]{2005MNRAS.363.1349G} Gierli\'{n}ski, M., \& Zdziarski, A.~A.\ 2005, \mnras, 363, 1349
\bibitem[Gierli{\'n}ski et al.(2008)]{2008MNRAS.388..753G} Gierli{\'n}ski, M., Done, C., \& Page, K.\ 2008, \mnras, 388, 753 

\bibitem[Gilfanov et al.(1999)]{1999A&A...352..182G}Gilfanov, M., Churazov, E., \& Revnivtsev, M.\ 1999, \aap, 352, 182

\bibitem[Gilfanov et al.(2000)]{2000Gil} Gilfanov, M., Churazov, E., \& Revnivtsev, M.\ 2000, \mnras, 316, 923 
\bibitem[Gou et al.(2011)]{2011Gou}Gou, L, et al., 2011, \apj, 742, 85 

\bibitem[Haardt et al.(1991)]{Haardt1991}Haardt, F., Maraschi, L., 1991, \apj, 380, L51-L54

\bibitem[Homan \& Belloni(2005)]{2005Ap&SS.300..107H} Homan, J., \& Belloni, T.\ 2005, \apss, 300, 107 

\bibitem[Ibragimov et al.(2005)]{2005MNRAS.362.1435I}Ibragimov, A., Poutanen, J., Gilfanov, M., Zdziarski, A.~A., \& Shrader, C.~R.\ 2005, \mnras, 362, 1435 

\bibitem[Ichimaru(1977)]{1977ApJ...214..840I} Ichimaru, S.\ 1977, \apj, 214, 840 

\bibitem[Ishisaki et al.(2007)]{xissim}Ishisaki, Y., et al., 2007, PASJ, 59, 113

\bibitem[Kawabata \& Mineshige(2010)]{KM2010} Kawabata, R., \& Mineshige, S.\ 2010, \pasj, 62, 621 
\bibitem[Kubota \& Done(2004)]{2004MNRAS.353..980K} Kubota, A., \& Done, C.\ 2004, \mnras, 353, 980 


\bibitem[Kokubun et al.(2007)]{Kokubun2007} Kokubun, M., et al.\ 2007, \pasj, 59, 53 

\bibitem[Koyama et al.(2007)]{Koyama2007} Koyama, K., et al.\ 2007, \pasj, 59, 23

\bibitem[Kubota et al.(2010)]{Kubota2010} Kubota, A., Done, C., Davis, S.~W., et al.\ 2010, \apj, 714, 860 

\bibitem[Kitamoto et al.(1984)]{Kitamoto1984} Kitamoto, S., Miyamoto, S., Tanaka, Y., et al.\ 1984, \pasj, 36, 731 

\bibitem[Kotov et al.(2001)]{2001MNRAS.327..799K} Kotov, O., Churazov, E., \& Gilfanov, M.\ 2001, \mnras, 327, 799 

\bibitem[Liang \& Price(1977)]{1977ApJ...218..247L} Liang, E.~P.~T., \& Price, R.~H.\ 1977, \apj, 218, 247 


\bibitem[Li et al.(2005)]{Li2005} Li, L.-X., Zimmerman, E.~R., Narayan, R., \& McClintock, J.~E.\ 2005, \apjs, 157, 335 

\bibitem[Magdziarz et al. (1998)]{1998M} Magdziarz, P., Blaes, M. O., Zdziarski, A.~A., Johnson, W. N., \& Smith, D. A., 1998, \mnras, 301, 179 

\bibitem[Makishima et al.(1986)]{1986ApJ...308..635M} Makishima, K., Maejima, Y., Mitsuda, K., et al.\ 1986, \apj, 308, 635 

\bibitem[Makishima et al.(2008)]{2008PASJ...60..585M} Makishima, K., et al.\ 2008, \pasj, 60, 585 

\bibitem[Miller et al.(2012)]{M2012} Miller, J.~M., Pooley, G.~G., Fabian, A.~C., et al.\ 2012, \apj, 757, 11 

\bibitem[Mitsuda et al.(1984)]{1984PASJ...36..741M}Mitsuda, K., et al.\ 1984, \pasj, 36, 741 

\bibitem[Mitsuda et al.(2007)]{Mitsuda2007} Mitsuda, K., et al.\ 2007, \pasj, 59, 1 

\bibitem[Miyamoto \& Kitamoto(1989)]{1989Natur.342..773M}Miyamoto, S., \& Kitamoto, S.\ 1989, \nat, 342, 773 

\bibitem[Miyamoto et al.(1992)]{1992ApJ...391L..21M}Miyamoto, S., Kitamoto, S., Iga, S., Negoro, H., \& Terada, K.\ 1992, \apjl, 391, L21 

\bibitem[Mehdipour et al.(2011)]{Mehdipour2011}Mehdipour, M., et al.\ 2011, \aap, 534, A39 

\bibitem[Narayan \& Yi(1995)]{1995ApJ...444..231N}Narayan, R., \& Yi, I.\ 1995, \apj, 444, 231 

\bibitem[Negoro et al.(1995)]{Negoro1995}Negoro, H., Kitamoto, S., Takeuchi, M., and Mineshige, S., 1995, 452, L49

\bibitem[Noda et al.(2011a)]{Noda2011a}Noda, H., et al. 2011a, \pasj, 63, 449

\bibitem[Noda et al.(2011b)]{Noda2011b}Noda, H., et al. 2011b, \pasj, 63, 925

\bibitem[Noda et al.(2013)]{2013PASJ...65....4N}Noda, H., et al.\ 2013, \pasj, 65, 4 

\bibitem[Nowak et al.(1999)]{1999ApJ...510..874N}Nowak, M.~A., Vaughan, B.~A., Wilms, J., Dove, J.~B., \& Begelman, M.~C.\ 1999, \apj, 510, 874 

\bibitem[Nowak et al.(2011)]{2011ApJ...728...13N} Nowak, M.~A., et al.\ 2011, \apj, 728, 13 

\bibitem[Oda et al.(1971)]{Oda1971} Oda, M., Gorenstein, P., Gursky, H., et al.\ 1971, \apjl, 166, L1 

\bibitem[Orosz and Hauschildt (2000)]{Orosz2000} Orosz, J. A., \& Hauschildt, P. H. 2000, A\&A, 364, 265
\bibitem[Orosz et al.(2011)]{Orosz2011} Orosz, J.~A., McClintock, J.~E., Aufdenberg, J.~P., et al.\ 2011, \apj, 742, 84 

\bibitem[Poutanen et al.(1993)]{1993Po} 
Poutanen, J., Vilhu, O., 1993, A\&A, 275, 337

\bibitem[Poutanen \& Svensson(1996)]{1996ApJ...470..249P}
Poutanen, J., \& Svensson, R.\ 1996, \apj, 470, 249 

\bibitem[Poutanen et al.(2008)]{Poutanen2008} Poutanen, J., Zdziarski, A.~A., \& Ibragimov, A.\ 2008, \mnras, 389, 1427 

\bibitem[Revnivtsev et al.(1999)]{2000Rev} Revnivtsev, M., Gilfanov, M., \& Churazov, E.\ 1999, \aap, 347, L23 

\bibitem[Reid et al.(2011)]{Reid2011} Reid, M.~J., McClintock, J.~E., Narayan, R., et al.\ 2011, \apj, 742, 83 

\bibitem[Remillard \& McClintock(2006)]{2006ARA&A..44...49R} Remillard, R.~A., \& McClintock, J.~E.\ 2006, \araa, 44, 49 

\bibitem[Ross \& Fabian(2005)]{RF2005} Ross, R.~R., \& Fabian, A.~C.\ 2005, \mnras, 358, 211 

\bibitem[Serlemitsos et al.(2007)]{2007PASJ...59S...9S} Serlemitsos, P.~J., et al.\ 2007, \pasj, 59, 9 

\bibitem[Shakura \& Sunyaev(1973)]{1973A&A....24..337S} 
Shakura, N.~I., \& Sunyaev, R.~A.\ 1973, \aap, 24, 337 

\bibitem[Shapiro et al.(1976)]{1976ApJ...204..187S} 
Shapiro, S.~L., Lightman, A.~P., \& Eardley, D.~M.\ 1976, \apj, 204, 187 

\bibitem[Shidatsu et al.(2011)]{Shidatsu2011} Shidatsu, M., et al.\ 2011, \pasj, 63, 785 

\bibitem[Sobolewska et al.(2011)]{Sobolewska2011} Sobolewska, M.~A., Papadakis, I.~E., Done, C., \& Malzac, J.\ 2011, \mnras, 417, 280 

\bibitem[Sunyaev \& Tr\"{u}mper(1979)]{1979Natur.279..506S}Sunyaev, R.~A., \& Tr\"{u}mper, J.\ 1979, \nat, 279, 506 

\bibitem[Takahashi et al.(2001)]{T2001} Takahashi, K., Inoue, H., \& Dotani, T.\ 2001, \pasj, 53, 1171 

\bibitem[Takahashi et al.(2008)]{2008PASJ...60S..69T}Takahashi, H., et al.\ 2008, \pasj, 60, 69

\bibitem[Takahashi et al.(2007)]{TadHXD} Takahashi, T., et al.\ 2007, \pasj, 59, 35

\bibitem[Takahashi et al.(2010)]{2010SPIE.7732E..27T} Takahashi, T., et al.\ 2010, \procspie, 7732,  

\bibitem[Tananbaum et al.(1972)]{Tananbaum} Tananbaum, H., Gursky, H., Kellogg, E., Giacconi, R., \& Jones, C.\ 1972, \apjl, 177, L5 

\bibitem[Thorne \& Price(1975)]{Thorne1975} Thorne, K.~S., \& Price, R.~H.\ 1975, \apjl, 195, L101

\bibitem[Torii et al.(2011)]{2011Torii} Torii, S., Yamada, S., Makishima, K., et al.\ 2011, \pasj, 63, 771 



\bibitem[Uchiyama et al.(2008)]{Uchiyama2008} Uchiyama, Y., et al.\ 2008, \pasj, 60, 35 


\bibitem[Uttley et al.(2011)]{Utt2011} Uttley, P., Wilkinson, T., Cassatella, P., et al.\ 2011, \mnras, 414, L60 


\bibitem[Veledina et al.(2011a)]{2011MNRAS.414.3330V} Veledina, A., Vurm, I., \& Poutanen, J.\ 2011a, \mnras, 414, 3330 
\bibitem[Veledina et al.(2011b)]{2011ApJ...737L..17V} Veledina, A., Poutanen, J., \& Vurm, I.\ 2011b, \apjl, 737, L17 

\bibitem[Wang et al.(2012)]{Wang2012} Wang, J.-M., Cheng, C., \& Li, Y.-R.\ 2012, \apj, 748, 147 


\bibitem[Wilkinson \& Uttley(2009)]{W2009} Wilkinson, T., \& Uttley, P.\ 2009, \mnras, 397, 666 

\bibitem[Xiang et al.(2011)]{2011Xiang} Xiang, J., Lee, J.~C., Nowak, M.~A., \& Wilms, J.\ 2011, \apj, 738, 78 

\bibitem[Yamada et al.(2009)]{2009GX339}Yamada, S.  et al., 2009, \apj, 707L, 109

\bibitem[Yamada(2011)]{2011PhD} Yamada, S.\ 2011, Ph.D.~Thesis, 

\bibitem[Yamada et al.(2011)]{2011GSO} Yamada, S., et al.\ 2011, \pasj, 63, 645 

\bibitem[Yamada et al.(2012)]{2012PL} Yamada, S., et al.\ 2012, \pasj, 64, 53 

\bibitem[Yamaoka et al.(2005)]{2005ChJAS...5..273Y} Yamaoka, K., Uzawa, M., Arai, M., Yamazaki, T., \& Yoshida, A.\ 2005, Chinese Journal of Astronomy and Astrophysics Supplement, 5, 273 

\bibitem[Yuan et al.(2007)]{2007Yuan} 
Yuan, F., Zdziarski, A.~A., Xue, Y., Wu, X., 2007, \apj, 659, 541 

\bibitem[Wardzi{\'n}ski, et al.(2000)]{WZ2000} Wardzi{\'n}ski, G., Zdziarski, A. A., 2000, MNRAS, 314, 183


\bibitem[Zdziarski et al.(1998)]{1998Zd} Zdziarski, A.~A., Poutanen, J., Mikolajewska, J., et al.\ 1998, \mnras, 301, 435 

\bibitem[Zdziarski et al.(2002)]{Z2002} Zdziarski, A.~A., Poutanen, J., Paciesas, W.~S., \& Wen, L.\ 2002, \apj, 578, 357 

\bibitem[Zdziarski \& Gierli{\'n}ski(2004)]{2004PThPS.155...99Z}Zdziarski, A.~A., \& Gierli{\'n}ski, M.\ 2004, Progress of Theoretical Physics Supplement, 155, 99 


\bibitem[Zwart et al.(2008)]{Zwart2008} Zwart, J.~T.~L., Barker, R.~W., Biddulph, P., et al.\ 2008, \mnras, 391, 1545 


\end{thebibliography}
\end{document}